\documentclass[%
twocolumn,
amsmath,
prx,
]{revtex4-2}

\usepackage{comment}
\usepackage{graphicx}
\usepackage{dcolumn}
\usepackage{bm}
\usepackage{amsfonts}
\usepackage{color}
\usepackage{multirow}
\usepackage{hhline}

\newcommand{\floor}[1]{\lfloor #1 \rfloor}

\UseRawInputEncoding

\begin{document}

\title{Fermion-to-qubit encodings with arbitrary code distance}

\author{Manuel G. Algaba$^{1,2}$}
\email{manuel.algaba@meetiqm.com}
\author{Miha Papi\v c$^{1,3}$}
\author{In\'es de Vega$^{1,3}$}
\author{Alessio Calzona$^{1}$}
\author{Fedor \v{S}imkovic IV$^{1}$}
\email{fedor.simkovic@meetiqm.com}

\affiliation{$^{1}$IQM Quantum Computers, Georg-Brauchle-Ring 23-25, 80992 Munich, Germany}
\affiliation{$^{2}$PhD Programme in Condensed Matter Physics, Nanoscience and Biophysics, Doctoral School, Universidad Aut{\'o}noma de Madrid}
\affiliation{$^{3}$Department of Physics and Arnold Sommerfeld Center for Theoretical Physics, Ludwig-Maximilians-Universit{\"a}t M{\"u}nchen, Theresienstr. 37, 80333 Munich, Germany}

\date{\today}

\begin{abstract}
We introduce a framework which allows to systematically and arbitrarily scale the code distance of local fermion-to-qubit encodings in one and two dimensions without growing the weights of stabilizers. This is achieved by embedding low-distance encodings into the surface code in the form of topological defects. We introduce a family of \emph{Ladder Encodings} (LE), which is optimal in the sense that the code distance is equal to the weights of density and nearest-neighbor hopping operators of a one-dimensional Fermi-Hubbard model.
In two dimensions, we show how to scale the code distance of LE as well as other low-distance encodings such as Verstraete-Cirac and Derby-Klassen. We further introduce \emph{Perforated Encodings}, which locally encode two fermionic spin modes  within the same surface code structure. We show that our strategy is also extendable to other topological codes by explicitly embedding the LE into a 6.6.6 color code. 
\end{abstract}

\maketitle

\section{Introduction}
Quantum computers are expected to excel at the simulation of fermionic systems, encompassing applications from quantum chemistry, solid-state and high-energy physics~\cite{McArdle_2020}. To this end, it is required to express fermionic Hamiltonians in the language of qubit operators. This can be achieved by either accounting for the fermionic anticommutation relations in the wavefunction~\cite{Su_2021} or in the operators~\cite{Jordan1928}. In the latter case, a fermion-to-qubit encoding is used to map the fermionic creation and annihilation operators to Paulis. Research focused on finding ideal transformations between fermions and spins dates back nearly 100 years to the seminal work of Jordan and Wigner~\cite{Jordan1928}. It has since experienced a renaissance in the past two decades, driven by the development of advanced numerical methods for the evaluation of fermionic problems on classical and quantum computers~\cite{Kitaev1995, Friesner2005}. 

One of the shortcomings of the original Jordan-Wigner transformation (JWT) is that it results in non-local Pauli operators in two and higher dimensional systems, whose weight grows with the system size. A number of \emph{local} fermion-to-qubit encodings have been designed to avoid this operator non-locality, albeit at the price of potentially introducing non-locality in states \cite{Guaita_2025}. Specifically, the operator locality preservation is enforced through the addition of ancillary qubits that enlarge the Hilbert space thus creating a redundancy in the quantum information. This redundancy manifests itself in the form of stabilizer operators which, in turn, can serve as a tool for quantum noise mitigation~\cite{chien2023}, error correction~\cite{landahl2023}, and also entanglement reduction in tensor network based simulations \cite{Parella2024}. Indeed, it has been shown that fermion-to-qubit encodings can be interpreted as quantum error correcting codes with potentially non-trivial logical code distances~\cite{Jiang2019}.

For quantum error correction (QEC) codes, as the number of available qubits grows, one generally has the choice between increasing the code distance, or alternatively encoding more logical qubits using multiple copies of a given QEC patch. In contrast, for the vast majority of fermionic encodings \cite{Derby2020,Verstraete2005,Ball2005,Chiew2021,Miller_2023,Jiang2020,Vlasov_2022,Setia2019} one can grow the size of the encoded fermionic system, but the only general way of increasing the logical distance is through concatenation, which usually results in high-weight stabilizers and large qubit counts. Alternatively, one can resort to exponentially scaling brute-force numerical search algorithms to search for fermionic encodings with higher code distance~\cite{Chen_2024, Chien2022, Simkovic2024}. This strategy, however, quickly becomes computationally prohibitive and higher-distance encodings found this way generally suffer from growing weights of stabilizers and logical operators, as well as growing connectivity requirements between qubits, making their hardware implementations largely unrealistic.

Recently, a scalable family of fermion-to-qubit encodings based on topological defects in the surface code was introduced as an efficient alternative to fault-tolerant computation based on the direct encoding of logical qubits \cite{landahl2023}. One major advantage on this approach is that increasing the code distance keeps the weights of all stabilizers constant. It is of interest to study, whether this construction can be generalised to arbitrarily increase the distance of the numerous low-weight and low-distance encodings found in literature~\cite{Verstraete2005, Derby2020, Setia2018,Chen_2023, Algaba_2024}. Another question is whether this is of practical use in pre-fault-tolerant quantum computational settings. This question boils down to the usefulness of encodings for quantum error mitigation (QEM) and particual quantum error correction (QEC) purposes. Recent benchmarks \cite{chien2023} have indicated that, at least for some noise rate settings, low-distance encodings perform better than both the JWT and their high-distance counterparts when used in combination with QEM methods based on stabilizer post-selection. However, it is unclear whether this trend persists to arbitrary encodings and quantum noise settings.

In this paper, we first introduce a one-dimensional fermion-to-qubit encoding with code distance $d=2$ and show how this distance can be arbitrarily grown by embedding the encoding into a surface code structure. We then proceed to generalise our construction to two dimensions and establish a connection to codes built from topological defects within the surface code, such as those presented in Ref.~\cite{landahl2023}. We show that, by expressing encodings in this language, one can equally generate families of encodings from other known low-distance encodings as starting points such as the Verstaete-Cirac encoding~\cite{Verstraete2005}, the Derby-Klassen compact encoding~\cite{Derby2020} and the hexagonal encoding from Ref.~\cite{Chien2022}. We further introduce an encoding family tailored for the study of fermionic Hamiltonians with two spin types. We show that our encodings can be extended using other topological codes while preserving the topological structure of the Majorana operators. Finally, we give insights on the effect of growing the distance when performing QEM and QEC.

The paper is structured as follows: We provide a brief introduction to the methodology of local fermion-to-qubit encodings in Section~\ref{subsec:local_encodings} and introduce the family of one-dimensional \emph{Ladder Encodings} in Sections~\ref{subsec:le_1d} and \ref{subsec:le_1d_growing}, giving their topological interpretation in Section~\ref{subsec:le_1d_topo}. In Section~\ref{subsec:le_2d}, we extend the LE family to the two-dimensional case and generalise the construction to other scalable encoding families in Section~\ref{subsec:others_2d}. In Section~\ref{section:perforated_encoding}, we introduce the \emph{Perforated Encoding} for two spin species. In Section~\ref{sec:color_code} we embed the LE into the 6.6.6 color code. Finally, we discuss the performance of high-distance encodings in the context of quantum error mitigation and quantum error correction in Section~\ref{sec:qem}.

\section{Edge-Vertex formalism}\label{subsec:local_encodings}
Fermionic Hamiltonians are typically constructed using the fermionic creation ($c_{i\sigma}^\dagger$), annihilation ($c_{i\sigma}$) and number operators ($n_{i\sigma} \equiv c_{i\sigma}^\dagger c^{\phantom{\dagger}}_{i\sigma}$), for a given spin $\sigma$ and lattice site $i$. Specifically, we will consider the Fermi-Hubbard model (FHM) (defined either on a chain or on a square lattice) in this work. The Hamiltonian reads:
\begin{align}
\mathcal{H}_{\text{FH}} & =  -  \sum_{i,j, \sigma} t^{ij} c^\dagger_{i \sigma}c_{j\sigma}^{\phantom{\dagger}}+U \sum_{i} n_{i\uparrow}^{\phantom{\dagger}}n_{i\downarrow}^{\phantom{\dagger}}.
\label{eq:hubbard}
\end{align} 

The above fermionic operators can be described by the edge and vertex formalism, which is an intermediate representation between Dirac fermionic operators ($c$, $c^\dagger$) and Majorana fermionic operators ($\gamma_i$,$\bar{\gamma}_i$)\cite{Bravyi2002}, which are themselves related by $\gamma_i = c^\dagger_i +c_i$ and $\bar{\gamma}_i = i(c^\dagger_i -c_i)$. Let us define the vertex $V_{i}=-i\gamma_i\bar{\gamma}_i$ and edge $E_{ij}=-i\gamma_i\gamma_j$ operators for every fermionic mode $i$ and pair of modes $(i,j)$ in terms of creating and annihilation operators:
\begin{align}
E_{ij}&=-i(c_ic_j+c^\dagger_ic^\dagger_j+c_ic^\dagger_j+c^\dagger_ic_j) \\
V_i&=c_ic^\dagger_i-c^\dagger_ic_i. \label{eq:vertex}
\end{align}
One can show that the vertex and edge operators obey the following commutation relations:
\begin{align}\label{eq:cond1} &\quad \quad \{E_{ij},V_i\}=\{E_{ij},E_{jk}\}=0 \\
 &[E_{ij},E_{kl}]=[E_{ij},V_{k}]=[V_i,V_j]=0 \nonumber
\end{align}
for indices $i\neq j\neq k \neq l$. Additionally, for edge operators the identities $E_{ji}=-E_{ij}$ and $E_{ik}=iE_{ij}E_{jk}$ exist.

The Hamiltonian terms from Eq.~\ref{eq:hubbard} can be expressed in terms of $E$ and $V$:
\begin{align}
    n_j^{\phantom{\dagger}} \rightarrow & \frac{1}{2}(1-V_j) \\
    c^\dagger_j c_k^{\phantom{\dagger}} + c^\dagger_k c_j^{\phantom{\dagger}} \rightarrow & \frac{i}{2}(V_k-V_j)E_{jk}.
    \label{eq:quadraticsumhc}
\end{align}

A popular approach in defining a fermion-to-qubit encoding is to proceed by picking a graph defined in terms of edges that connect vertices, which correspond to the fermionic modes of the Hamiltonian to be simulated \cite{Bringewatt_2023}. The particular choice of graph structure will have an influence on the properties of the encoding and once chosen, edges and vertices are identified with tensor products of Pauli operators acting on qubits, which have to obey the correct commutation relations as dictated by Eq.~\ref{eq:cond1}. Instead of identifying Pauli strings with edge operators, we can directly consider the two terms on the r.h.s. of Eq.~\ref{eq:quadraticsumhc}, which we call transfer operators:
\begin{align}\label{eq:transferoperator}
    T_{ij} \equiv \frac{i}{2}V_iE_{ij}  \quad\quad
    T_{ji} \equiv -\frac{i}{2}V_jE_{ij}
\end{align}

and which obey the following commutation relations:
\begin{align}
 \{ T_{ij}, V_i\} =& \{T_{ij},T_{kj}\} = 0 \\
 [T_{ij},V_k] = [T_{ij}, T_{kl}] =& [T_{ij},T_{ji}] = [T_{ij},T_{jk}] = 0. \nonumber
\end{align}
with $i<j<k$. It is easily seen that every closed loop of edges defines a stabilizer of the encoding, which commutes with all logical operators. The distance $d$ of an encoding is defined by the lowest-weight of a logical operator in the encoding, and it is possible to detect all errors of weight up to $d-1$ and correct all errors up to weight $\floor{(d-1)/2}$ \cite{Nielsen2012}.

Here, we will consider the one-dimensional chain and the two-dimensional square graphs for a single spin species, with the exceptions of Section~\ref{section:perforated_encoding} where an encoding for two spin types is presented. Generalizing any fermion-to-qubit encoding to models with more than one spin type can also be done by assigning alternating fermionic modes in one lattice direction to different spins or by superposing multiple lattices so that each lattice encodes the modes of a single spin species. The latter construction preserves the code distance of the encoding, since any undetected error would have to commute with the stabilizers of both lattices which requires a Pauli string of weight at least $d$. This construction enables the implementation of all logical operators of the FHM.

\begin{figure}
    \centering
    \includegraphics[width=\linewidth]{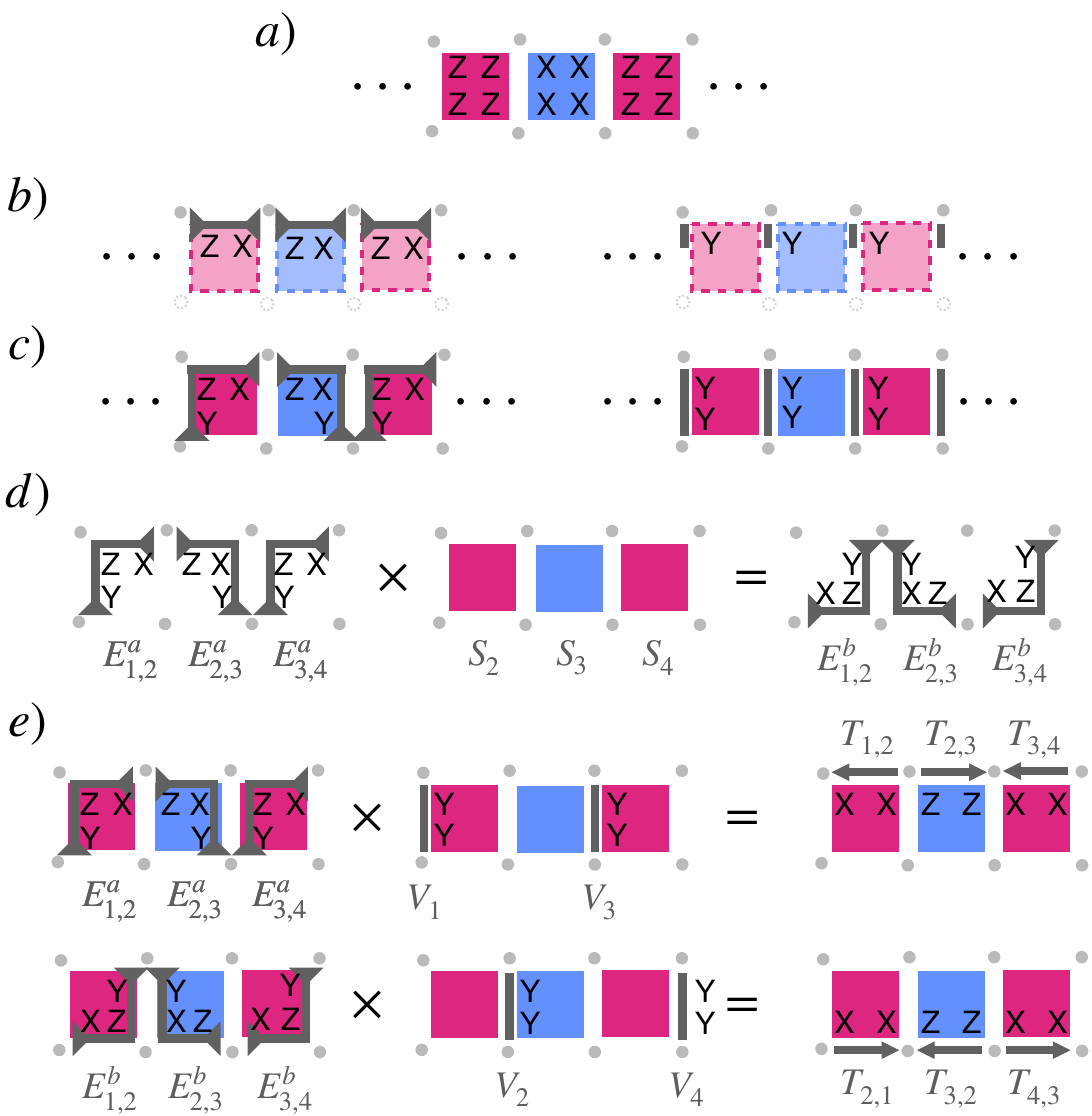}
    \caption{\textbf{Understanding the Ladder Encoding}. The LE can be obtained by taking a) surface code stabilizers: $Z$ plaquettes (red) and $X$ plaquettes (blue) and b) embedding the Jordan-Wigner Transformation in the surface code: edge operators (inwards-pointing arrows) and vertex operators (straight lines). However, the logical operators anticommute with stabilizers if embedded directly, which can be solved by c) acting on additional qubits with $Y$ Paulis so the commutation relations are fixed and distance $d=2$ is ensured. d) Logical equivalence up to stabilizer multiplication between two different types of edge operators, $E^a$ and $E^b$. This allows for the parallelization of e) the two transfer operators $T_{ij}$ and $T_{ji}$ (arrows) which are constructed from $E^a$ and $E^b$.}
    \label{fig:jw_to_le}
\end{figure}

\section{One-dimensional Ladder Encodings}\label{sec:1d}

\subsection{LE$(d=2)$}\label{subsec:le_1d}

In Fig.~\ref{fig:jw_to_le} we show a graphical construction of what we will call the Ladder Encoding (LE), starting from the widely adopted Jordan-Wigner Transformation (JWT). To construct a higher-distance encoding from the JWT we attempt to embed it into the top row of qubits of a surface code with alternating stabilizers of $\bar{Z} = Z_j^a Z_j^b Z_{j+1}^a Z_{j+1}^b$ and $\bar{X} = X_{j+1}^a X_{j+1}^b X_{j+2}^a X_{j+2}^b$ type, depicted as red and blue squares in Fig.~\ref{fig:jw_to_le}a, respectively. Our starting point, the JWT, has a one-dimensional edge-vertex graph and encodes every fermionic mode with a single qubit, as shown in Fig.~\ref{fig:jw_to_le}.b. In the JWT, we define a fermionic creation operator as:
\begin{align} \label{eq:JWT}
c^{\dagger}_j = \frac{1}{2} Y_1 Y_2 \dots Y_{j-1} (X_j - i Z_j),
\end{align}
where $j$ is an index within a chosen ordering that accounts for the fermionic lattice position and spin. We note that this is a slightly different definition to that generally found in literature as we have exchanged the Pauli $Y$ and $Z$ operators' roles. This, however, will have no effect on the properties of the encoding. Eq.~\ref{eq:JWT} leads to the following definitions for edge and vertex operators in the JWT (shown as gray operators on the left and right sides of Fig.~\ref{fig:jw_to_le}b, respectively):
\begin{align}
    V_j = \frac{1}{2}(1-Y_j) \\
    E_{j,j+1} = Z_j X_{j+1}.
\end{align}

\begin{figure*}[ht]
    \centering
    \includegraphics[width=0.9\linewidth]{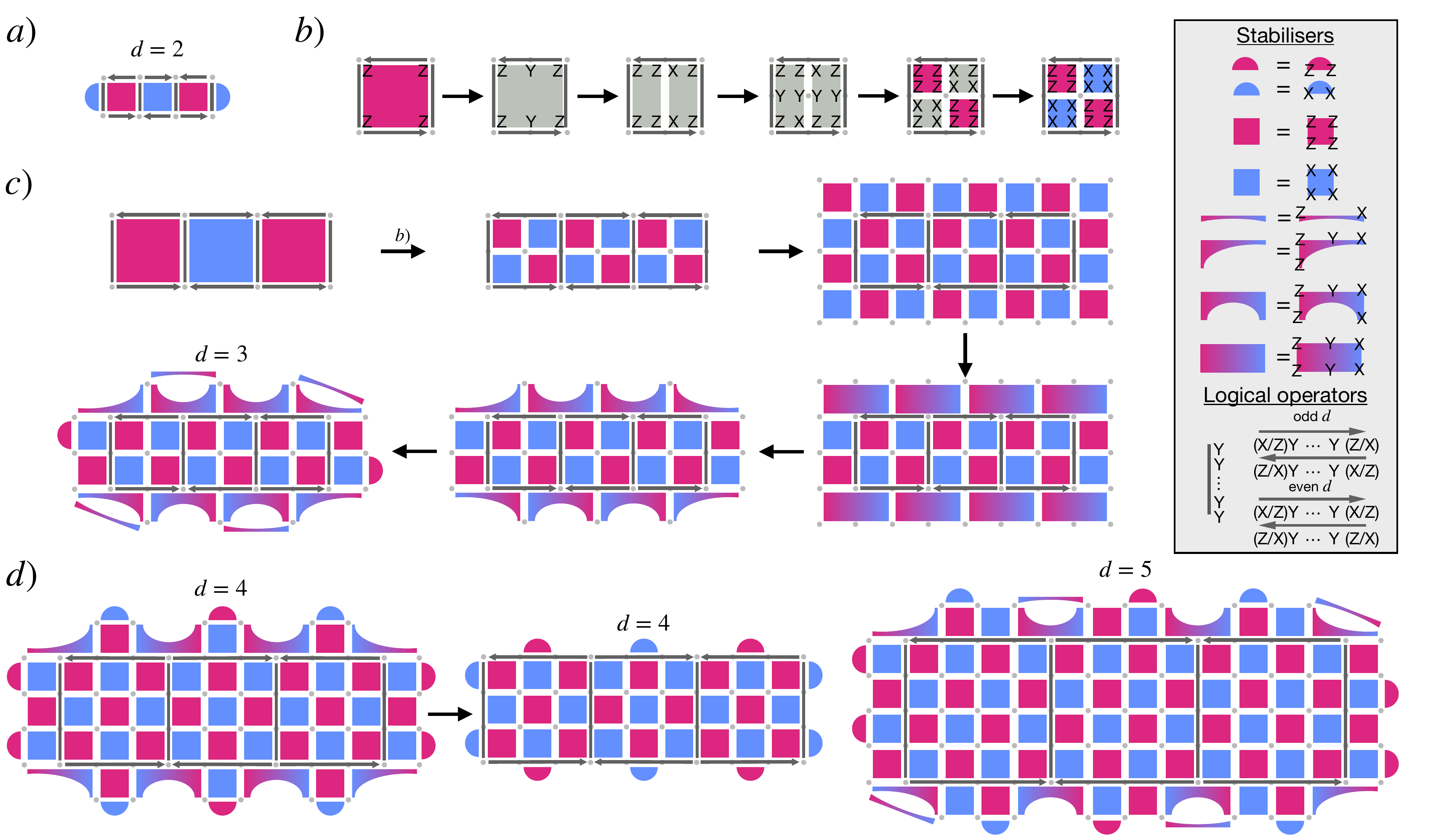}
    \caption{\textbf{Structure of one-dimensional Ladder Encodings}. a) LE$(d=2)$ four fermionic modes, b) successive decomposition of a surface code stabilizer into four surface code stabilizers by adding extra qubits, c) strategy for growing the distance of the Ladder Encoding from LE$(2)$ to LE$(d)$, d) Ladder Encodings LE$(d=\{4,5\})$. Note that for even distances the boundary structure can be simplified, as shown for LE$(d=4)$.}
    \label{fig:le_distances}
\end{figure*}

The distance of the JWT is defined by the weight of its lowest-weight logical operator, in this case the vertex, with $d=1$. If we consider the JWT on a closed chain rather than an open string, we can identify the only stabilizer present in the encoding which is a loop containing all edges of the JWT. This stabilizer has support on all qubits and is of the form: $S_{\text{JW}} = \bigotimes_j Y_j$.

In order to increase the logical distance of the encoding we have to at least 1) introduce additional qubits to enlarge the Hilbert space which will create additional stabilizers that can be used to identify errors and 2) increase the weights of all logical operators to be equal to or larger than the desired code distance. If we naively attempt to embed the JWT within the surface code as suggested above, this results in the logical operators anticommuting with the stabilizers of the surface code. In order to fix these commutation relations, we grow the vertex and edge operators by making them act with a Pauli-$Y$ operator on the qubits in the bottom row (see Fig.~\ref{fig:jw_to_le}c). This construction now satisfies the correct commutation relations and guarantees that the code distance is $d=2$, given that it is embedded in a surface code of that distance.

For each vertex pair, it is possible to identify two weight-three edges, as shown in Fig.~\ref{fig:jw_to_le}d, which are logically equivalent and related to each other by multiplication with a stabilizer defined on the in-between plaquette. Remarkably, this combination of edges and vertices results in two weight-two transfer operators acting on distinct rows of qubits on the ladder (see Fig.~\ref{fig:jw_to_le}e), which concludes the intuitive derivation of the new encoding.

Let us now proceed to define what we will call the one-dimensional Ladder Encoding with distance 2, LE$(d=2)$. It consists of the assignment of the vertex and transfer operators, defined on a one-dimensional chain graph as follows:
\begin{align}
    V_i &= Y_i^a Y_i^b \\
    T_{i,i+1} &=
    \begin{cases}
        X_i^a X_{i+1}^a & \text{for}\; i \in \text{even} \nonumber\\
        Z_i^a Z_{i+1}^a  & \text{for}\; i \in \text{odd}\nonumber\\
    \end{cases} \\
    T_{i+1,i} &=
    \begin{cases}
         X_i^b X_{i+1}^b & \text{for} \;i \in \text{even} \nonumber\\
         Z_i^b Z_{i+1}^b & \text{for} \;i \in \text{odd}\nonumber \\
    \end{cases}
\end{align}
where indices $a$ and $b$ correspond to the two sides of the $i$-th ladder rung. This choice satisfies all commutation relations between transfer and vertex operators.

Additionally, we define a group of stabilizers $\mathcal{S}=\{S_i\}^{m+1}_{i=1}$ where $m$ is the number of fermionic modes and
\begin{align}
S_i  = 
\begin{cases} 
      Z_i^a Z_i^b Z_{i+1}^a Z_{i+1}^b & \text{if} \ 1<i<m, \; i \in \text{even} \\
      X_i^a X_i^b X_{i+1}^a X_{i+1}^b & \text{if} \ 1<i<m, \; i \in \text{odd}. \\
\end{cases}
\label{eq:stabs}
\end{align}
At the beginning and end of the ladder, one must additionally define two-qubit (digon) stabilizers:
\begin{align}
\label{eq:self-loop}
S_{1} &= X_1^a X_1^b \\
S_m  &= 
\begin{cases} 
      X_{m+1}^a X_{m+1}^b & \text{if} \ m \in \text{even} \\
      Z_{m+1}^a Z_{m+1}^b & \text{if} \ m \in \text{odd}. \\
\end{cases} \nonumber
\end{align}
It can be easily checked that the distance of the code defined by the choice of stabilizers and logical operators is $d=2$ as every single-qubit error will anticommute with at least one stabilizer.
It is common in fermion-to-qubit encodings to identify stabilizers as products of closed loops of edge operators~\cite{Derby2020,Algaba_2024}. Given that we have defined an encoding with a one-dimensional connectivity graph, it should, intuitively, not contain any closed loops of edges forming stabilizers. However, it can be noticed that by applying Eq.~\ref{eq:transferoperator}, two different edge operators can be obtained from one another (as shown in Fig.\ref{fig:jw_to_le}d).

One can interpret this pair of edges as forming a loop between the same two vertices, thus generating a stabilizer:
\begin{equation}
    S_i = E_{i,i+1}^{a} E_{i,i+1}^{b}.
\end{equation}
Similarly, one could interpret the weight-two stabilizers at the respective ends of the ladder, and defined in Eq.~\ref{eq:self-loop}, as \emph{self-loops} consisting of single edge operators.

\subsection{Higher-distance one-dimensional Ladder Encodings LE$(d)$}\label{subsec:le_1d_growing}
Let us now describe a protocol for growing the code distance of the LE$(d=2)$ encoding, shown in Fig.~\ref{fig:le_distances}a, to obtain any higher-distance LE$(d)$ encoding. The strategy consists of splitting the weight-four surface-code stabilizers into two by introducing a pair of qubits along a given axis, and letting the two resulting stabilizers act on these new qubits with Pauli $X$ and $Z$  operators, respectively. Simultaneously, we extend the logical operators to act with Pauli $Y$ operators on all qubits which have been introduced along their original path. Finally, one needs to make sure that stabilizers which share a qubit over the diagonal act on it with the same Pauli operator (second-to-last step in Fig.~\ref{fig:le_distances}b). These requirement can always be satisfied and all stabilizers can collectively be rearranged (up to a trivial sign prefactor) into one of the $\bar{X}$ or $\bar{Z}$ forms of Eq.~\ref{eq:stabs} (last step of Fig.~\ref{fig:le_distances}b). 
This ensures that the mutual commutation relations of all logical operators stay unchanged whilst increasing the distance in the bulk of the encoding.

Upon inspection, one can observe that low-weight undetected errors can still occur at the boundaries of resulting encodings, which are tackled by the following steps. Consider the example of LE$(d=3)$ encoding four fermionic modes, as shown in Fig.~\ref{fig:le_distances}c. Here, the first step is identical to Fig.~\ref{fig:le_distances}b to increase the distance in the bulk of the encoding. The second step in Fig.~\ref{fig:le_distances}c consists of padding the lattice with one layer of surface code stabilizers on the outside. As a result, some of the new stabilizers will not commute with the logical operators, but this can be resolved by merging neighboring pairs of stabilizers, which would individually anticommute with a given logical operator at the top/bottom ends of the encoding (third step in Fig.~\ref{fig:le_distances}c). In the fourth step, all boundary qubits which are acted on by only a single stabilizer, and are thus redundant for encoding purposes, are removed. At this stage, undetectable two-qubit errors may still appear at the boundaries of the lattice, so the final step consists of adding selected two-qubit stabilizers to catch these errors. Following this procedure, all higher-distance encodings can be obtained in a similar fashion. Examples with four fermionic modes for LE($d=\{4,5\}$) are shown in Fig.~\ref{fig:le_distances}d. We point out that, in the case of the one-dimensional LE, the operator weights are optimal in the sense that all vertex and transfer operator have a weight, which is equal to the code distance $d$.

A simpler construction, which maintains the distance of the encoding, exists for even distances with $d\geq4$, as shown in Fig.~\ref{fig:le_distances}d. It requires  only the addition of weight-two stabilisers on the boundaries and does not introduce any additional qubits to the encoding. These configurations coincide with the $(2n+2)-on$ Majorana codes, of which the tetron and hexon codes are a subfamily, which encode two and three fermionic modes, respectively~\cite{Litinski_2018}. We note that this simplification is not possible for odd distances without increasing the weight of logical operators in the bulk to $d+1$. 

\begin{figure}
    \centering
    \includegraphics[width=0.8\linewidth]{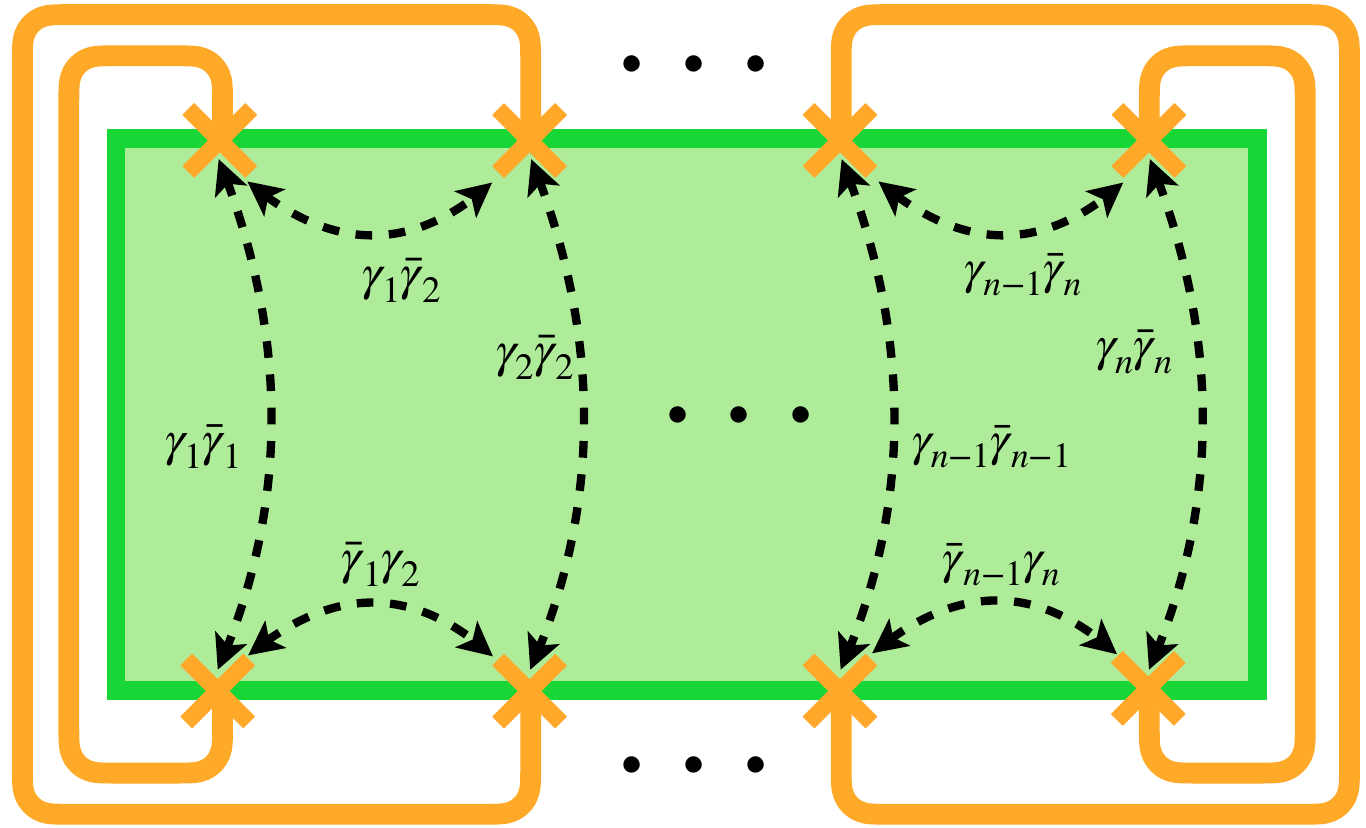}
    \caption{\textbf{Topological structure of the 1D LE$(d)$ encoding}. Connected yellow crosses represent pairs of twist defects at the boundary of the encoding (green) and black dashed arrows are associated with quadratic Majorana operators.}
    \label{fig:le_1d_top}
\end{figure}

\subsection{Topological interpretation}\label{subsec:le_1d_topo}

Whilst we have derived the above encodings using the formalism of edge and vertex operators, they can alternatively be described by the language of topological defects within the surface code. Indeed, topological error correcting codes are closely related to Majorana fermions and thus can be used to generate fermion-to-qubit encodings \cite{brown2017}.

In the surface code, if all stabilizers (including the external blank space) can be colored such that no two neighboring stabilizers have \textit{i)} the same color and \textit{ii)} no more than one shared side, then it is said to be 2-\textit{colorable} and contains no Majorana operators. This is, i.e. the case for a surface code with only a single smooth boundary (a boundary made up of X stabilizers).
However, a surface code does not need to be 2-\textit{colorable} and the number of points in which the colorability breaks, also called colorability defects, is proportional to the number of Majorana operators in the code \cite{bombin2010}. In particular, pairs of Majorana operators are created by twist defects, which is topologically equivalent to having a transition between smooth (X-stabilizer) and rough (Z-stabilizers) boundaries \cite{brown2017}. 

In this regard, any LE$(d)$ encoding can be seen as an explicit fermion-to-qubit encoding that arises from the introduction of twist defects in the surface code too. A topologically equivalent structure to all one-dimensional LE$(d)$ encodings is shown in Fig.~\ref{fig:le_1d_top}. Here, twist defects are placed at the vertical boundaries to generate pairs of Majorana operators, and each yellow cross corresponds to a colorability defect introduced by a weight-five stabilizer of the one-dimensional LE$(d)$ encoding. Each defect can be assigned a Majorana operator and logical operators involving products of neighboring pairs of Majorana operators (as shown in Fig.~\ref{fig:le_1d_top}) are equivalent to the vertex and transfer operators ($V_i$, $T_{ij}$, $T_{ji}$) we have previously defined in Eqs.~\ref{eq:vertex} and \ref{eq:transferoperator}.

\section{Two-dimensional Ladder Encodings}\label{sec:2d}

\subsection{2D LE$(d)$}\label{subsec:le_2d}

\begin{figure*}[ht]
    \centering
    \includegraphics[width=0.8\linewidth]{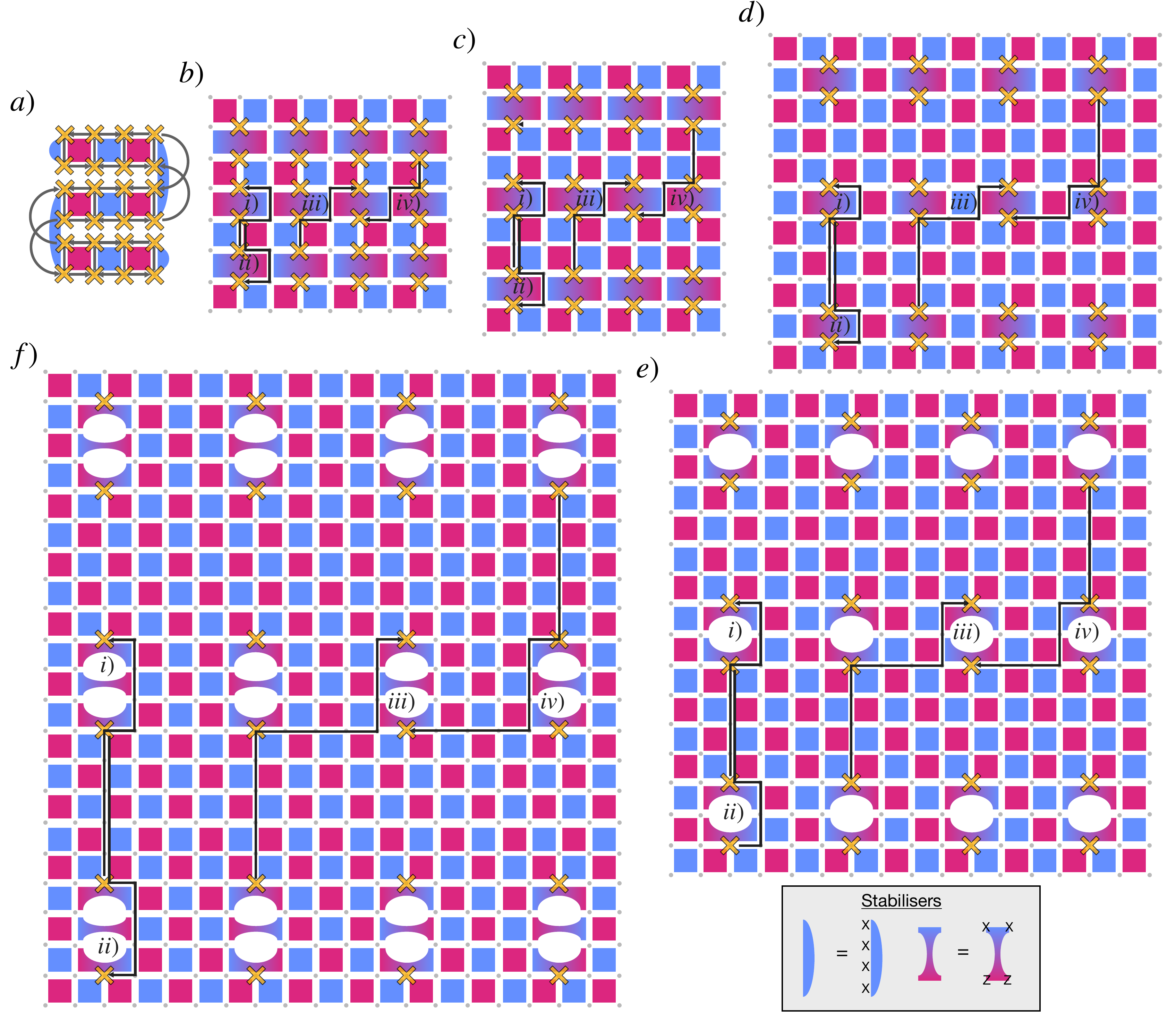}
    \caption{\textbf{Structure of the 2D LE$(d)$ encodings}. a) Non-local snake-like pattern for connecting multiple rows of LE$(d)$. Local 2D LE$(d)$ for b) $d=2$, c) $d=3$, d) $d=4$, e) $d=5$ and f) $d=6$ showing vertical transfer operators $i)$, $ii)$ and diagonal transfer operators $iii)$, $iv)$.}
    \label{fig:le_2d}
\end{figure*}

Let us now turn our attention to the generalization of Ladder Encodings LE$(d)$ to two dimensions. The most straightforward approach is similar in spirit to the Jordan-Wigner encoding, which is to vertically stack multiple rows of the 1D LE$(d)$ in a snake-like pattern as shown in Fig.~\ref{fig:le_2d}a. Even though this method works for ladder encodings of any distance, it does not preserve the locality of all logical operators for two-dimensional fermionic Hamiltonians, thus yielding a non-scalable version of the LE that may only be beneficial for small system sizes.

\begin{figure*}[ht]
    \centering
    \includegraphics[width=0.8\linewidth]{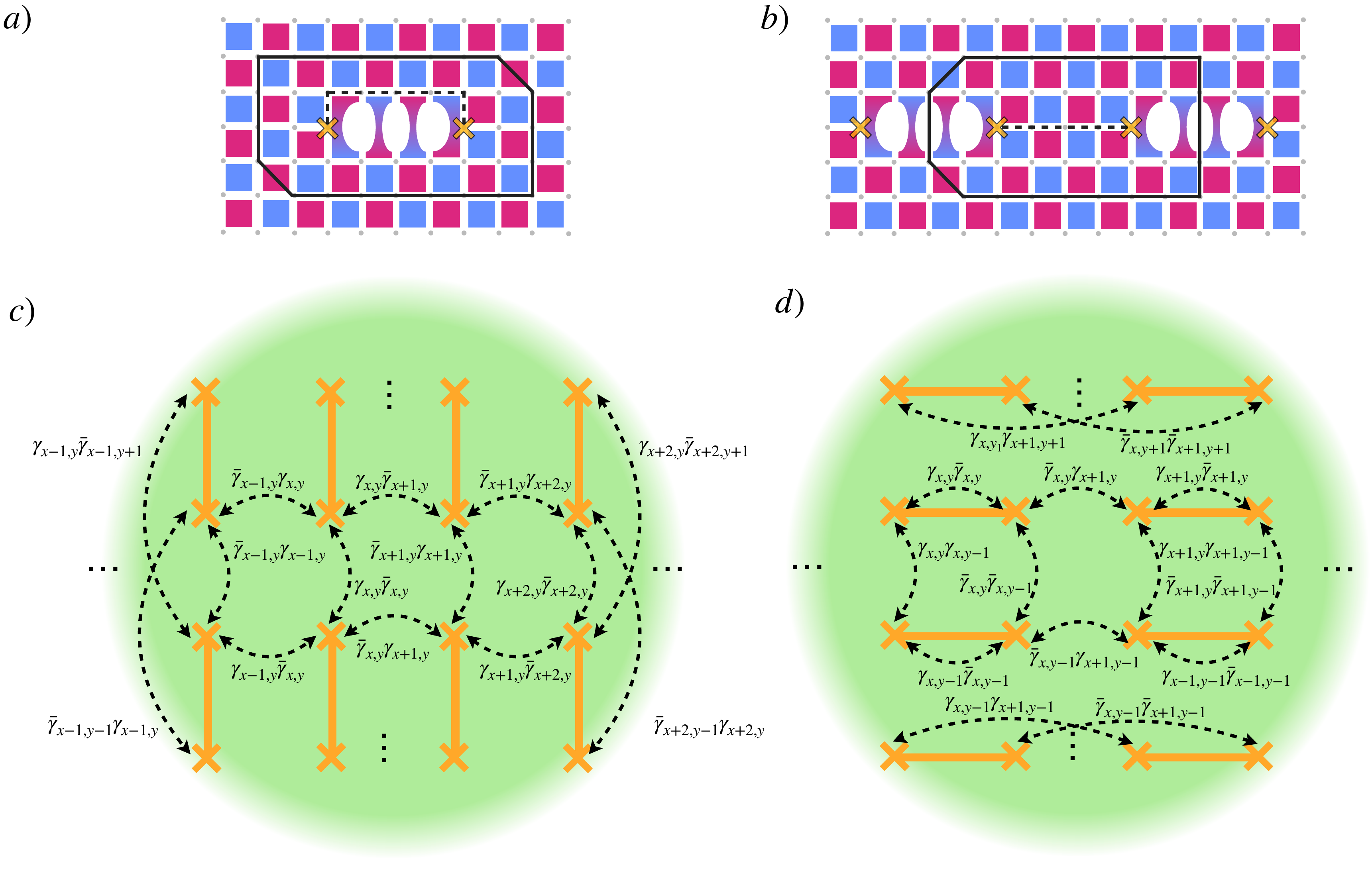}
    \caption{\textbf{Topological structure of the 2D LE$(d)$ encodings.} Equivalence between loop-like logical operators (solid) and their line version with reduced weight (dashed) for a) loops around twist defects and b) loops intersecting twist defects. c) Majorana ordering of LE$(d)$  and d) Majorana ordering of the encoding from Ref.~\cite{landahl2023}.}
    \label{fig:le_topological}
\end{figure*}

Alternatively, it is possible to use the fact that for  LE$(d\geq2)$ our construction introduces weight-six stabilizers, such as those shown in Fig.~\ref{fig:le_distances}c, each of which gives rise to a pair of Majorana fermions. These can, in turn, be used to extend the LE pattern vertically, as shown in Fig.~\ref{fig:le_2d}b. In order to introduce vertical (or diagonal) transfer operators it is sufficient to implement Pauli strings between the defects of the corresponding Majorana operators in a way that obeys the correct commutation relations with stabilizers and other logical operators. It is straightforward to verify that these are indeed satisfied for the encoding of Fig.~\ref{fig:le_2d}b and, given that all logical operators have weight of at least two, we get $d=2$ for the encoding. Two-dimensional encodings for LE$(d = \{3,4,5,6\})$, as well as their vertical and diagonal transfer operators, are show in Fig.~\ref{fig:le_2d}b-e. We note that, for LE$(d = 5)$ and LE$(d = 6)$, we can no longer use single weight-six stabilizers, since weight-four logical operators exists between the two twist-defects they generate, which would limit the distance of encodings to $d\leq4$. We must therefore split the weight-six stabilizers into two weight-five stabilizers for LE$(d = 5)$, growing the weight of the logical operator to five, and we can move the two defects even further apart for LE$(d = 6)$ by introducing an intermediate irregular weight-four stabilizer. These defects can be arbitrarily pulled  apart to achieve even higher code distances by introducing additional irregular weight-four stabilizers to separate them further. The weights of vertex and horizontal transfer operators for these encodings are equal to $d$, identical to the case of one-dimensional LE$(d)$ encodings. At the same time, all vertical transfer operators are of weight $2d-1$, or alternatively the they can be evenly split into weight $d$ and weight $3d-2$ operators. 

Note that, in this setup, multiplying the Pauli string that represents a logical operator by stabilizers gives rise to exponentially many (in the number of independent stabilizers) equivalent logical operators whose different weights can have an impact on the performance of a simulation using a given encoding. For example, consider the solid line in Fig.~\ref{fig:le_topological}a which is, in principle, a less desirable representation of the logical operator that connects a pair of twist defects compared to the reduced-weight version which is depicted by a dashed line. Further, note that the pairing of pairs of twist-defects can be done arbitrarily, but we use the term 'pair of twist defects' to refer to colorability defects directly connected by a group of irregular stabilizers. 

Not only the Pauli string chosen for each logical operator but also the ordering and assignment of the Majorana pairs that represent each fermionic mode has an impact on the performance of an encoding. In Ref.~\cite{landahl2023} a similar two-dimensional construction to LE$(d)$ using twist-defects in the surface has been introduced. The main difference to the LE (Fig.~\ref{fig:le_topological}c) is that for the encodings from Ref.~\cite{landahl2023}, whose Majorana representation in the bulk is depicted in Fig.~\ref{fig:le_topological}d,
different pairs of Majorana operators are assigned to vertex and transfer operators. This results in different weights for logical operators in Ref.~\cite{landahl2023}, where two out of four transfer operators in 2D are of optimal weight $d$, but density and density-density operators are of higher weight. The logical operators found in Ref.~\cite{landahl2023} are depicted as loops (black lines in Fig.~\ref{fig:le_topological}a and Fig.~\ref{fig:le_topological}b), but they can straightforwardly be reduced to their more compact low-weight representations (dashed lines). A full comparison of the two families of encodings can be found in Table~\ref{table:comparison}.

\begin{table*}
\centering
\begin{tabular}{|c|c|c|c|c|c|c|c|} 
\hline
\multicolumn{1}{|l}{} & \multicolumn{1}{l|}{} & \multicolumn{4}{c|}{This work}        & \multicolumn{2}{c|}{Previous work}  \\ 
\cline{3-8}
\multicolumn{1}{|l}{} &                       & LE$(d)$ & VC$(d)$ & HX$(d)$  & DK$(d)$ & LM$(d)$ & Red. LM$(d)$              \\ 
\hline \hline

\multirow{4}{*}{1D}   & $T^H_{i,i+1}$           & \multicolumn{2}{c|}{$d$}     & $d$     & $d$     & $2d+12$  & $d$                       \\ 
\cline{2-8}
                      & $T^H_{i+1,i}$           & \multicolumn{2}{c|}{$d$}    & $d$     & $d$     & $6d+4$ & $3d-2$                      \\ 
\cline{2-8}
                      & $n_i$                   & \multicolumn{2}{c|}{$d$}     & $d$     & $d$     & $2d+4$  & $d$                     \\ 
\cline{2-8}
                      & $n_in_{i+1}$            & \multicolumn{2}{c|}{$2d$}    & $2d$    & $2d$    & $4d+2$  & $2d$                    \\ 
\hline
\multirow{2}{*}{2D}   & $T^V_{i,i+1}$           & \multicolumn{2}{c|}{$2d-1$}   & $2d-1$   & $2d-1$   & $4d+2$ & $2d-1
$                       \\ 
\cline{2-8}
                      & $T^V_{i+i,i}$           & \multicolumn{2}{c|}{$2d-1$}    & $2d-1$    & $2d-1$    & $4d+2$ & $2d-1$                       \\
\cline{2-8} 
\hline \hline

\multirow{7}{*}{$r$}   & $d=1$           & -   & $2$    & -  & $1.5$      & \multicolumn{2}{c|}{-}                    \\ 
\cline{2-8}
                      & $d=2$           & \multicolumn{2}{c|}{$4$}     & $2$      & $4$       & \multicolumn{2}{c|}{-}                      \\ 
\cline{2-8}
                      & $d=3$                 & \multicolumn{2}{c|}{$6$}    & $4$    & $7.5$         & \multicolumn{2}{c|}{-}                     \\ 
\cline{2-8}
                      & $d=4$            & \multicolumn{2}{c|}{$12$}    & $9$    & $16$      & \multicolumn{2}{c|}{-}                   \\ 
\cline{2-8}
                      & $d=5$            & \multicolumn{2}{c|}{$23$}    & $15$    & $29.5$      & \multicolumn{2}{c|}{-}                   \\ 
\cline{2-8}
                      & $d=6$            & \multicolumn{2}{c|}{$38$}    & $28$    & $36$      & \multicolumn{2}{c|}{-}                   \\ 
\cline{2-8}
                      & $d \geq 7$            & \multicolumn{2}{c|}{$2d^2-7d+8$}   & $2d^2-9d+10$     & $2d^2-5d+4.5$     & \multicolumn{2}{c|}{$2d^2-7d+8$}                 \\ 
\hline
\end{tabular}
\caption{\textbf{Comparison of different encodings.} Logical operator weights corresponding to the two-dimensional single-spin Fermi-Hubbard model are listed between four different families arbitrary-distance encodings developed in this work and compared to the encoding family introduced in Ref.~\cite{landahl2023} (original and optimally reduced operator weights are shown). The fermion-to-qubit ratios $r$ are reported for distances up to four and for odd distances beyond $d\geq 7$.}
\label{table:comparison}
\end{table*}

Note that the identification of lines connecting defects with quadratic Majorana operators can be done due to the anyonic nature of the surface code, which hosts two types of anyons, corresponding to one of the possible independent X- or Z-lines defined on it. Under fusion (line multiplication) of the two different types, the anyons give rise to a fermionic algebra, allowing us to identify closed loops of mixed X- and Z-lines with quadratic Majorana operators. Lines that cross the defect line of a pair of twist defects, exchange their Pauli labels $X\leftrightarrow Z$, so that loops surrounding two defects from different pairs of defects can be constructed, half being an X-line and half being Z-line. These loops anticommute with any other loop of the same nature surrounding the same defect. These can be reduced by stabiliser multiplication to straight lines connecting defects, which allows identifying quadratic Majorana operators with straight lines.

\begin{figure}[ht!]
    \centering
    \includegraphics[width=0.99\linewidth]{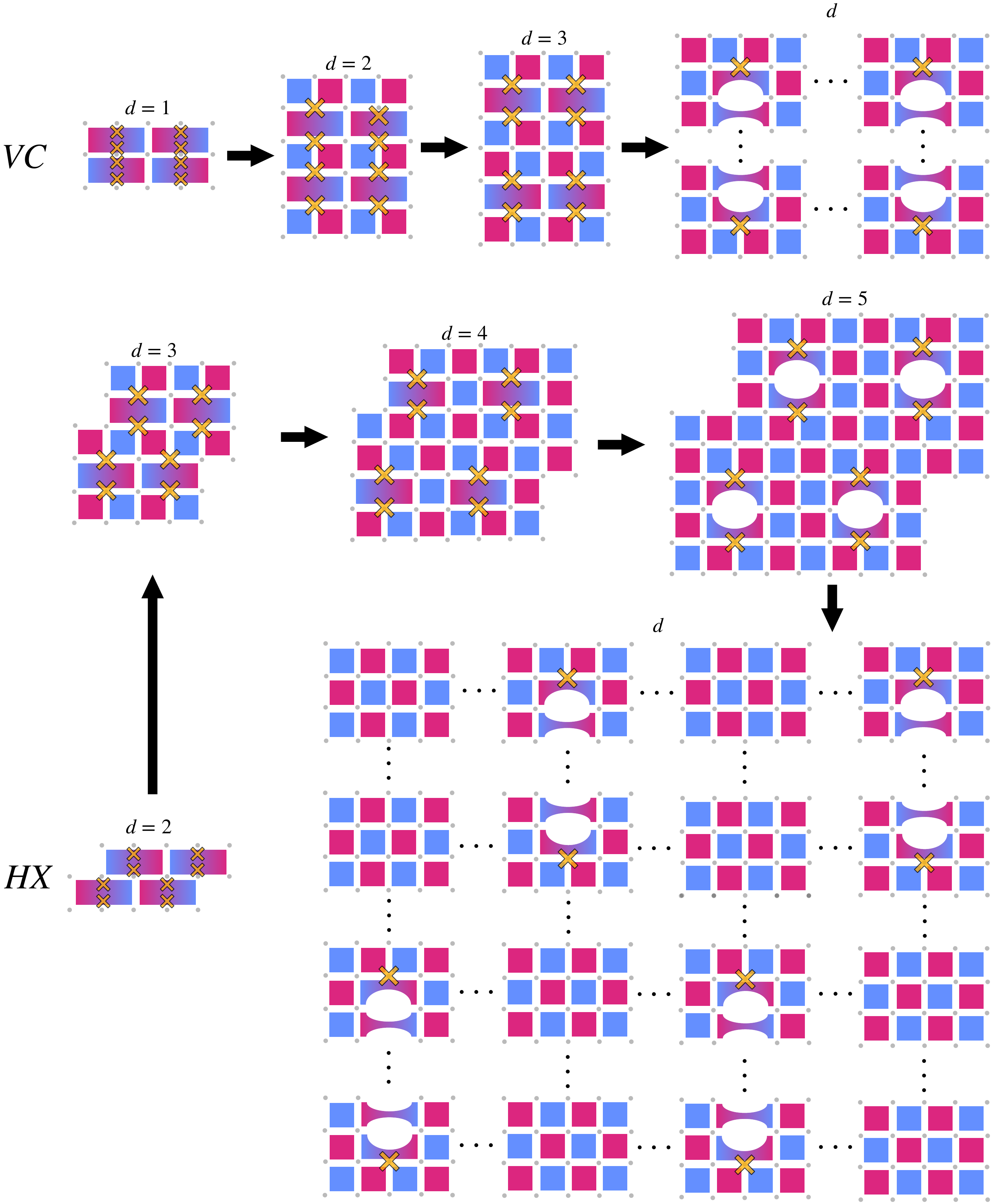}
    \caption{\textbf{Growing the distance of encodings from literature I}. Yellow crosses represent colorability defects that host Majorana operators and their separation establishes the code distance. Top: Verstraete-Cirac encoding \cite{Verstraete2005}, bottom: hexagonal encoding from Refs~\cite{Chien2022,Simkovic2024}.}
    \label{fig:other_encodings}
\end{figure}

\subsection{Growing the distance of other 2D encodings}\label{subsec:others_2d}
The strategy used to grow the distance of Ladder Encodings in previous sections can be equally adapted to other encodings. Specifically, in this section we use it to grow the distance of the Verstraete-Cirac (VC) encoding \cite{Verstraete2005}, the hexagonal encoding defined in Refs.~ \cite{Chien2022,Simkovic2024} and the compact encoding by Derby-Klassen (DK)~\cite{Derby2020}. It is possible to find an embedding of the stabilizers of all aforementioned encodings into a surface code-like structure as shown in Fig.~\ref{fig:other_encodings}. These embeddings contain colorability defects that host Majorana operators from which the higher-distance encodings are built. The separation between defects defines the minimal weight of logical operators (and logical errors). Consequently, separating these defects allows for an arbitrary increase in the distance of the underlying encodings, same as in the case of LE$(d)$. 

Let us focus on the original VC encoding from Ref.~\cite{Verstraete2005} shown at the top left of Fig.~\ref{fig:other_encodings}, which has distance $d=1$ limited by the weight-one vertex operators of the encoding. This encoding is stabilized by weight-six stabilizers, each of which hosts two colorability defects which are separated by four qubits along the stabilizer perimeter and defects from different neighboring stabilizers are horizontally three qubits apart. Furthermore, each qubit sandwiched by different two weight-six stabilizers supports two different colorability defects generated by them. This is another way of recognizing that weight-one logical operators exist in the encoding, which limits the distance of VC to $d=1$.

\begin{figure*}[ht!]
    \centering
    \includegraphics[width=0.8\linewidth]{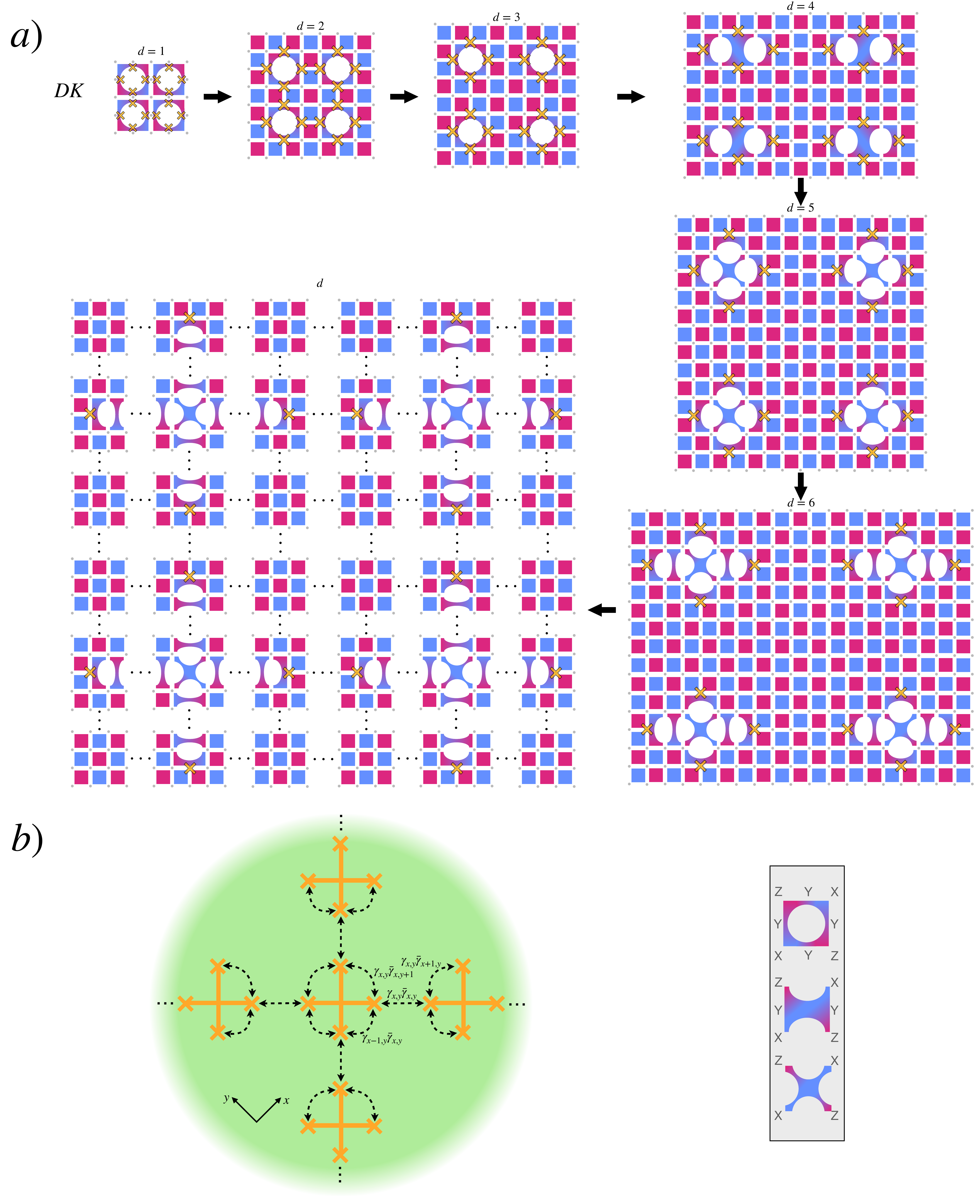}
    \caption{\textbf{Growing the distance of encodings from literature II}. a) Qubit representation of the Derby-Klassen compact encoding \cite{Derby2020} for a range of distances where yellow crosses represent colorability defects that host Majorana operators and their separation establishes the distance of each encoding. b) Topological interpretation of the DK encoding.}
    \label{fig:other_encodings_dk}
\end{figure*}

To grow the distance of the VC encoding, it is necessary to pull apart the defects from different weight-six stabilizers by introducing a layer of surface code weight-four stabilizers in between (see Fig.~\ref{fig:other_encodings}). This way, all defects are sufficiently separated and the minimal weight logical operators that can be found in the new encoding  is two, thus growing the distance to $d=2$. At this point, it becomes clear that the resulting encodings for distances $d\geq2$ are in fact equivalent to those we obtained for the two-dimensional LE. This is not surprising, since both encodings contain the same type and arrangement of (weight-six) colorability defects.

The main difference of the original HX encoding (also shown in Fig.~\ref{fig:other_encodings}) compared to VC$(d=1)$ is that its weight-six stabilizers are positioned with a horizontal offset, which increases the separation of the defects and results in the higher distance of $d=2$ for the HX encoding. In order to further grow the distance of the HX encodings, one proceeds in much the same way as for the VC encoding. As a matter of fact, the two-dimensional LE and VC encodings have the same topological structure as HX, and all three result in the same logical operator weights for any given distance. The main difference is the shifted horizontal position between alternating rows of twist-defects which, however, has an impact on the slightly improved fermion-to-qubit ratio for HX, especially for small distances.

Finally, let us consider the DK encoding (see Fig.~\ref{fig:other_encodings_dk}a), which contains weight-eight stabilizers that generate four twist-defects, double that of the weight-six stabilizers found in LE, VC and HX. Similarly to these encodings, the distance is grown by pulling apart the defects, which can topologically be viewed as combinations of horizontal and a vertical pairs of defects (see Fig.~\ref{fig:other_encodings_dk}b). The intersection of the two pairs may generate different weight-four to weight-six stabilizers, as shown for DK$(d\geq4)$ encodings in Fig.~\ref{fig:other_encodings_dk}a. Nevertheless, this family of encodings also results in the same weights of logical operators as LE, VC and HX, albeit at a slightly worse fermion-to-qubit ratio.

We note that our strategy of generating higher-distance encodings is not deterministic and different ways of splitting the stabilizers to separate defects can result in different higher-distance generalizations of the same encoding. Indeed, the only feature of the surface code which this approach relies on is that it is a topological quantum error correction code whose anyons can resemble fermionic algebras. This means that if an embedding of a fermionic encoding is found within a different topological code, this strategy can be straightforwardly be applied to obtain a different generalization of the same initial encoding. In Section~\ref{sec:color_code} we show how to build a family of encodings topologically equivalent to those of the 1D LE based on the 6.6.6 color code \cite{bombincolorcode2006} instead of the surface code.

\begin{figure}
    \centering
    \includegraphics[width=\linewidth]{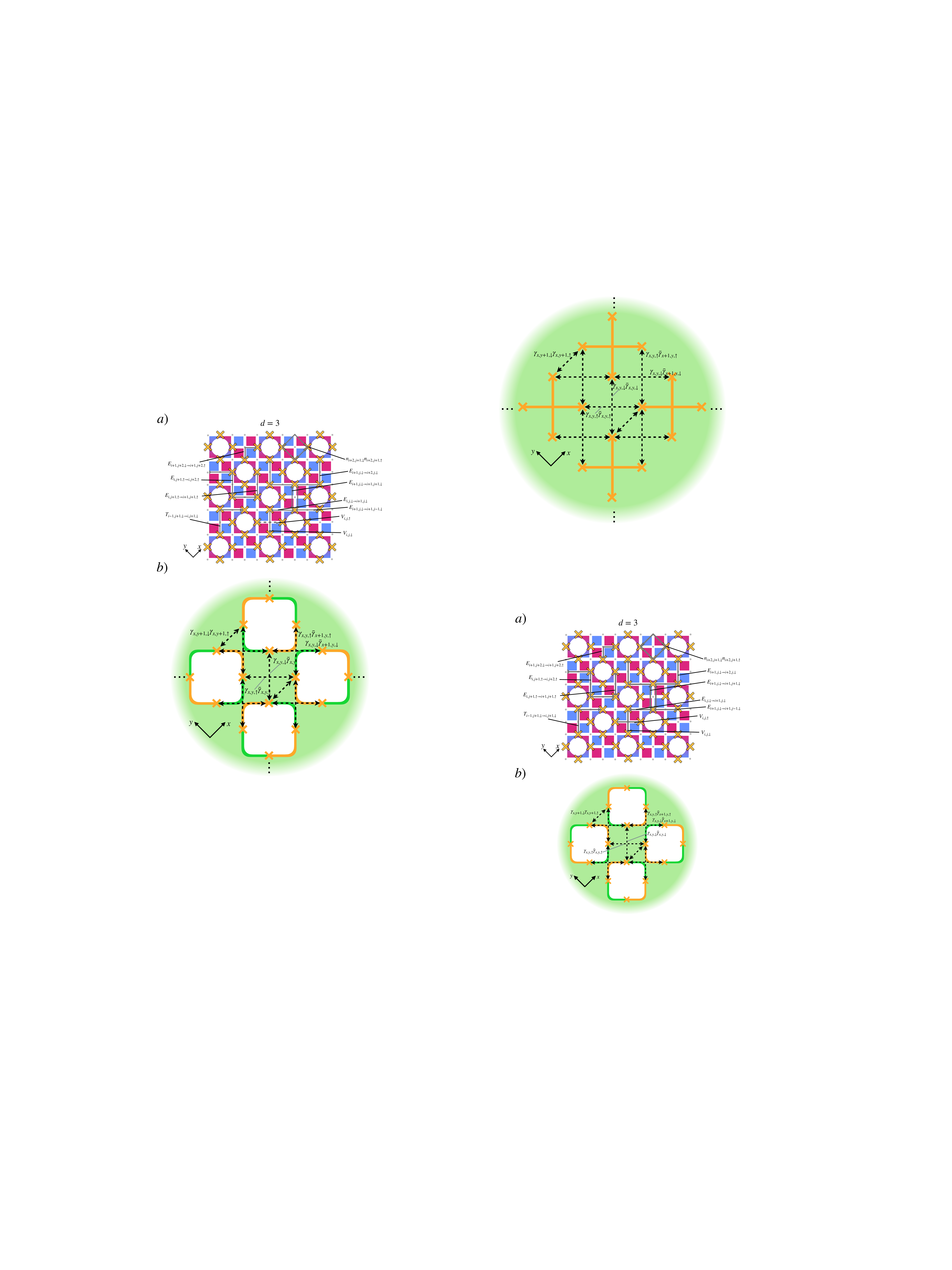}
    \caption{\textbf{The spinful Perforated Encoding}. a) Qubit representation of the PE$(d=3)$ where each group of $2\times 2$ plaquette stabilisers represents a fermionic lattice site encoding two spin modes. Vertex operators connect defects on two weight-eight stabilisers across a single site while the edge operators connect defects on different sites or defects with different spins. Note that all same-spin transfer operators between nearest neighbours are weight-five. b) Topological perspective of the PE.}
    \label{fig:pe_d}
\end{figure}

\section{Perforated Encodings for multi-spin lattice Hamiltonians}
\label{section:perforated_encoding}

So far, all encodings have been designed for fermionic systems with a single spin type. Possible ways of embedding spinful Hamiltonians into these encodings consist of either having multiple copies of an encoding, one for each spin type, or placing fermionic modes of different spins on the same lattice in an alternating manner. Both of these approaches, however, have their shortcomings. On the one hand, multiple copies of the same encoding may not fit into the square surface code qubit layout and connectivity, especially if qubits measuring stabilizers are also taken into account. Such encodings also do not allow for Hamiltonian terms to be defined which require edges between fermionic modes of different spin. On the other hand, naively embedding multiple spin species into the same encoding graph either leads to more non-local logical operators, or the requirement for fSWAP operations, both of which increase the circuit complexity.

\begin{figure}[b]    \centering\includegraphics[width=0.95\linewidth]{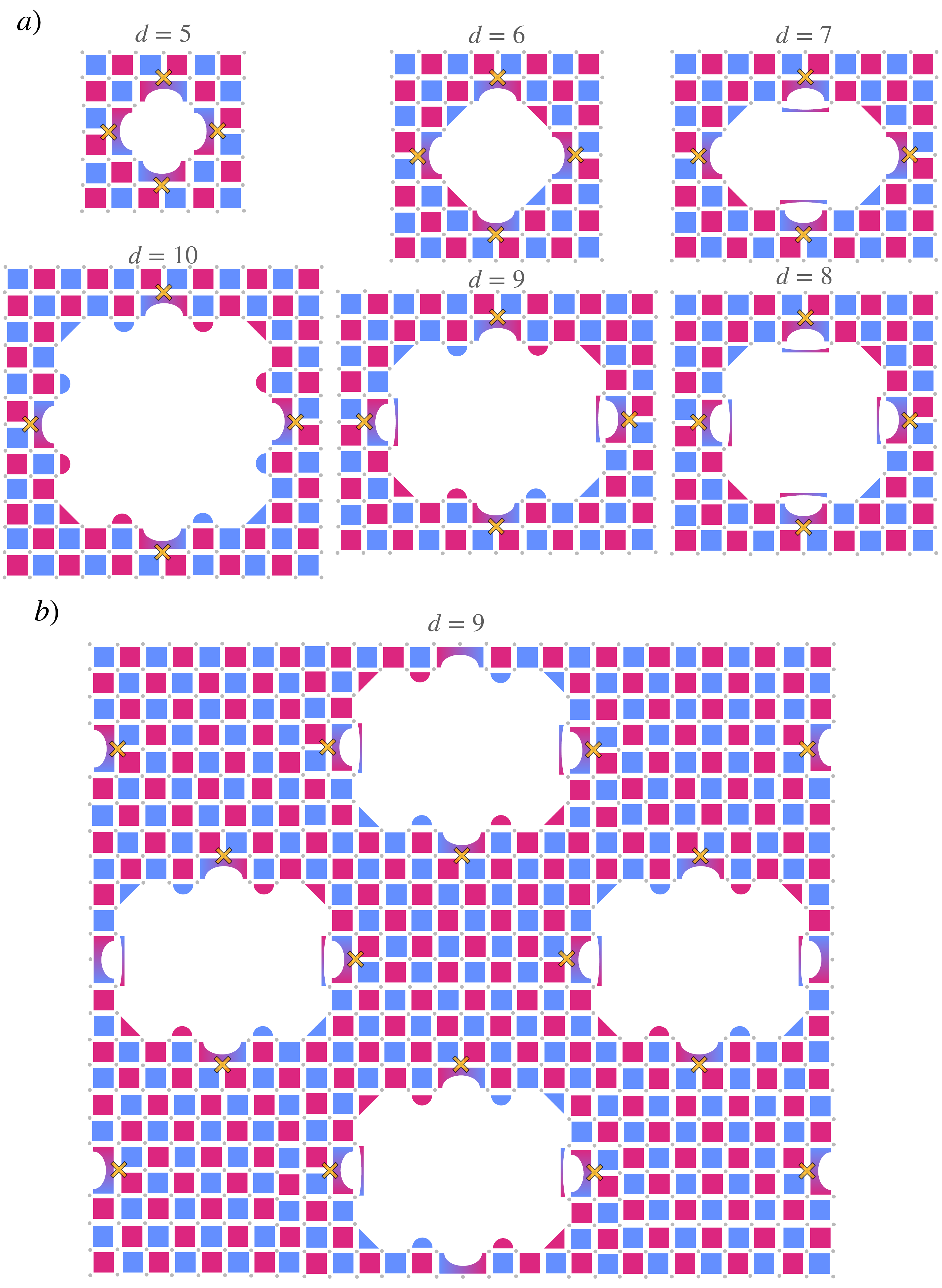}
    \caption{\textbf{Pulling apart defects inside holes.} Top: when growing the distance of a encoding, it is possible to substitute strings of higher-weight and non-local stabilizers  by holes with internal boundaries. Bottom: in some cases, the distance can be improved by placing holes with 
    alternating internal boundaries (rough vs. smooth) next to each other. An example for PE$(d=9)$ is shown.}
\label{fig:hole_defect}
\end{figure}

\begin{figure}
    \centering\includegraphics[width=0.75\linewidth]{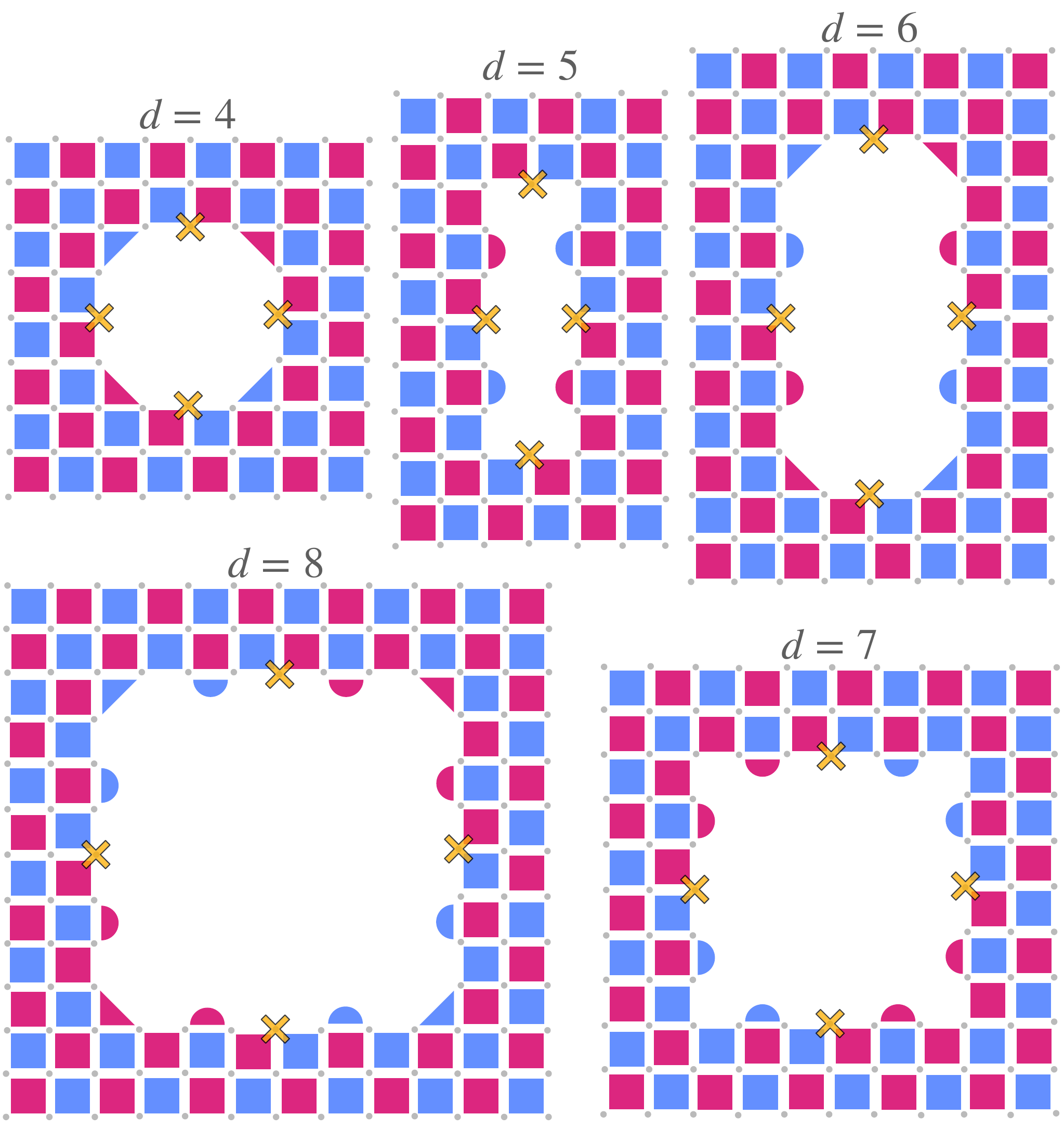}
    \caption{\textbf{Holes with low-weight stabilisers.} Holes of different distances for the Perforated Encoding using only weight-four and lower stabilisers.}
    \label{fig:low_weight_holes}
\end{figure}

Alternatively, the defect-based formalism of previous sections can also be used to directly derive efficient encodings for fermionic  Hamiltonians with multiple spin species, such as the spinful FHM, or multi-band FHM models. Similar to how the HX$(d)$ encoding can be topologically interpreted as a more compact version of the VC$(d)$ encoding with shifted alternating rows of stabilizers, we can also shift alternating rows of our DK$(d)$ encoding to build a compressed version of it, which we call the Perforated Encoding (PE). The distance $d=3$ Perforated Encoding which is shown in Fig.~\ref{fig:pe_d}a encodes locally two spin species within the same set of stabilizers. The vertices for two fermionic spin modes of a given lattice site correspond to horizontal and vertical weight-three operators within this encoding. Notably, this combination results in weight-four density-density terms, compared to weight-six for the single-spin encodings we have considered before. All edges are weight-three, while all transfer operators are weight-five, with the exception of spin-flip terms on involving modes of a single lattice site, which are also weight-three. 

To grow the distance of PE we need to decide on a way of breaking its stabilisers to separate defects further apart. Instead of deforming the stabilisers of PE$(d)$ as we have done for DK$(d)$, we can alternatively transform them into internal holes~\cite{brown2017}, with multiple types of smooth and rough boundaries along the internal perimeter. Holes with twist defects conforming to distance $5\leq d \leq 10$ are shown in Fig.~\ref{fig:hole_defect}. When generating encodings based on holes, it is important that diagonally neighboring holes have opposite colors on their closest boundaries (smooth vs. rough), as otherwise they would give rise to low-weight logical operators which would limit the distance of the encoding. The topological depiction of the PE encoding with twist-defects along internal boundaries is shown in Fig.~\ref{fig:pe_d}b.  Using holes has the added benefit of reducing the total number of required qubits and thus improving the fermion-to-qubit ratio of an encoding without otherwise changing its properties. We note that the holes used in this section, and shown in Fig.~\ref{fig:pe_d}a, are not unique and versions without weight-five stabilisers can often be identified, as shown in Fig.~\ref{fig:low_weight_holes}. This is a similar reduction to the one used in Subsection~\ref{subsec:le_1d_growing} for even distances, where for one-dimensional LE encodings $(2n+2)$-on codes could be used to eliminate the weight-five stabilisers.

In Table~\ref{table:weight_ops_pe} we summarize the weights of all logical operators and the fermion-to-qubit ratios for PE$(d)$. Crucially, this ratio is roughly four times better than for encodings from the previous section, since we have gained a factor-two by shifting weight-eight stabilizers from the DK encoding and another factor-two by using holes in the encoding. We note the concept of holes could equally be applied to the other two-dimensional encodings presented in the previous sections in order to improve their fermion-to-qubit ratios. 

\begin{table}[]
    \centering
    \begin{tabular}{|c|c|}
    \hline
        Operator & Pauli weight  \\ \hline 
        \hline
        $V_{i,j,\sigma}$ & $d$ \\ \hline
        $E_{i,j,\sigma\rightarrow i+1,j,\sigma}$ & $d\ ||\ 3d-2$\\ \hline 
        $E_{i,j,\sigma\rightarrow i,j+1,\sigma}$ & $d\ ||\ 3d-2$ \\ \hline 
        $E_{i,j,\sigma\rightarrow i,j,\bar{\sigma}}$ & $d$ \\ \hline
        \hline
        $n_{i,j,\sigma}$ & $d$ \\ \hline
        $n_{i,j,\sigma}n_{i,j,\bar{\sigma}}$ & $2d-1$ \\ \hline
        $n_{i,j,\sigma}n_{i+1,j,\bar{\sigma}}$ & $2d$ \\ \hline
        $T_{i,j,\sigma\rightarrow i+1,j,\sigma}$ & $2d-1$ \\ \hline
        $T_{i,j,\sigma\rightarrow i,j+1,\sigma}$ & $2d-1$ \\ \hline
        $T_{i,j,\sigma\rightarrow i+1,j+1,\sigma}$ & $3d-2 \ ||\ 4d-3$ \\ \hline
        \hline
        $r(d=3)$ & $3.5$ \\ \hline
         $r(d\geq 5)$ & $\frac{d^2}{2}+2d-6$ \\ \hline
    \end{tabular}
    \caption{\textbf{Operator weights fermion-to-qubit ratios for the Perforated Encoding.} Note that some operators might differ for odd and even sites, as indicated by vertical bars in the table. Fermion-to-qubit ratios for $d\geq5$ are valid uniquely for odd distances.}
    \label{table:weight_ops_pe}
\end{table}

\section{Fermionic encodings onto other topological codes}\label{sec:color_code}
The strategies for generating high distance encodings used along this paper are mainly based on the topological nature of the surface code as we have depicted in Figs.~\ref{fig:le_1d_top},\ref{fig:le_topological},\ref{fig:other_encodings_dk} and \ref{fig:pe_d}. Thus the methods shown can be applied to other topological error correction codes if the same topological structure is preserved.

\begin{figure}
    \centering 
    \includegraphics[width=\linewidth]{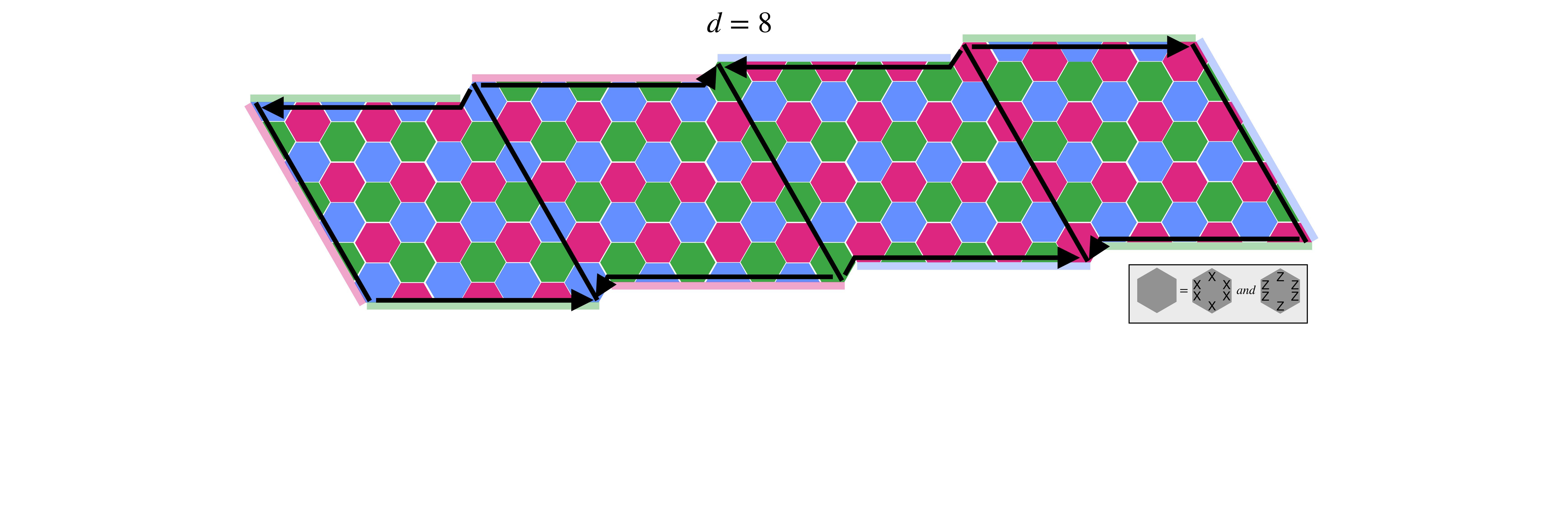}
    \caption{\textbf{$\mathrm{LE}(8)$ in the 6.6.6 color code.} In this code, each hexagon represents two stabilisers, an X and a Z plaquette. The vertical straight lines (vertex operators) are Y-Pauli strings and the horizontal lines are X-Pauli strings and Z-Pauli strings representing transfer operators.}
    \label{fig:color_code}
\end{figure}

To exemplify this statement  we embed the one-dimensional LE in a $d=8$, 6.6.6 color code \cite{bombincolorcode2006,landahl2011faulttolerantquantumcomputingcolor} in Fig.~\ref{fig:color_code}, but other color codes could also be used~\cite{Tiurev_2024}. This code consists of hexagonal plaquettes with two stabilisers at each plaquette, $XXXXXX$ and $ZZZZZZ$. Different instances of the color code can be chosen based on the specific boundaries \cite{Kesselring_2018}. Pauli-type boundaries add extra stabilisers on the boundaries to avoid errors on the boundaries, while color-type boundaries add no extra stabilisers but only cut existing ones into weight-four stabilisers.  We choose the boundaries to be color-type boundaries. Note that colors in this code do not represent different stabilisers as in the surface code but rather the nature of the anyons. Nine types of anyons can be identified in color codes \cite{Kesselring_2024} whose labels are chosen from the possible colors $\{r,g,b\}$ of the plaquettes that the anyon is crossing and the Pauli operators used $\{x,y,z\}$. Anyons commute with all the stabilisers except for the ones in the starting and ending points. This means that only anyonic excitations exist that connect defects on the lattice, such as the boundary color changes that we use in Fig.~\ref{fig:color_code}. They anticommute if they share no common label. In our example, we identify the vertices with $gy$, $ry$ and $by$, while for the left-headed arrows representing transfer operators we use $gx$, $rx$ and $bx$ and for the right-headed arrows representing transfer operators we use $gz$, $rz$ and $bz$.

In a surface code twist defect it is possible to create as many mutually anticommuting operators as desired, since one can draw a line across a defect to exchange the anyon type, giving rise to a fused anyon that anticommutes with all other fused anyons arising from that defect that only surround two defects. This makes extending the LE from 1D to 2D straightforward. However, in the color code version of the 1D LE we use only simple anyons and only three of them can mutually anticommute. Thus, for obtaining a 2D LE at least one extra anticommuting operator would be required and this can only be obtained by including fused color code anyons. For these, more complex defects would be necessary, like T- or B-defects \cite{Kesselring_2018}. This means that, even though these encodings can be generally extended to other codes, the simplicity of the 2D LE on the surface code is not necessarily achievable for arbitrary other topological codes.

\section{Consequences of growing the distance}\label{sec:qem}

Using the presented framework of constructing higher-distance local fermion-to-qubit encodings such as the 1D $\mathrm{LE}(d)$ to encode one-dimensional systems may seem pointless given that the Jordan-Wigner transformation is already local for such systems. However, the measurement of the resulting stabilizers allows to perform quantum noise mitigation or quantum error correction while keeping the fermionic operator weights as small as possible. Moreover, one-dimensional encodings can be a useful primitive for the study of fermionic systems with more complex connectivity graphs. For example, the implementation of electronic structure Hamiltonians from quantum chemistry requires simulating quartic terms of the form $c^\dagger_jc^\dagger_kc_l^{\phantom{\dagger}}c_m^{\phantom{\dagger}}$. One can show that these can be transformed into a sum of operators of the form $T_{jl}T_{km}$, which can be readily implemented within the LE as long as the fermionic modes corresponding to indices $j,k,l,m$ can be arranged to be neighbors on the fermionic chain graph. This is achieved through fSWAP operations, which can be implemented efficiently by using exponentials of vertex and transfer operators \cite{Kivlichan2018}. 

One downside of the higher-distance encodings derived here is that, compared to surface codes, they may require beyond-square lattice qubit connectivity. At a minimum, for the one-dimensional $\mathrm{LE}(d=2)$ encoding, data qubits require five connections, three to other data qubits and two to syndrome qubits which measure stabilizers. Simultaneously, each syndrome qubit requires four connections to data qubits contained in the corresponding stabilizer. In the case of a construction with two spin species, an additional data-data qubit connection is required. For higher-weight and two-dimensional encodings, the data qubit connectivity in the bulk grows to eight for one spin type, and nine for two spin types, which can be reduced back down to eight by using PE. All higher-distance encoding families we have presented here also require the measurement of higher-weight stabilizers with weights of up to eight. These stabilizers can either be implemented using a single syndrome qubit, or alternatively measured by multiple syndrome qubits located at positions corresponding to the surface code stabilizer tiling. For weight-five and weight-six stabilizers this requires two syndrome qubits while for weight-eight stabilizers four syndrome qubits must be used. One can then show that with the help of these additional syndrome qubits all higher-weight stabilizers can be measured within the standard surface code error correction cycle, albeit the code performance may differ. Especially at higher distances, the fact that the stabilizer weight and connectivity is independent on the distance presents a significant advantage compared to previous high-distance encodings found by brute force search methods \cite{Chen_2023, Simkovic2024}.

A natural question that arises from the previous sections is how the higher-distance encodings we have identified can benefit the simulation of fermionic systems on current and future quantum computers. Arguably, some of the most useful primitives in quantum simulation are evolution operators under single Hamiltonian terms $e^{-t\alpha_jh_j}$, where $H=\sum \alpha_j h_j$ is the Hamiltonian to be simulated. However, no transversal implementation of all such terms for an interacting fermionic Hamiltonian can be found given that 1) interacting fermionic Hamiltonians can be seen as matchgates between nearest and next-nearest neighbours which form a universal gate set \cite{Projansky2024} and 2) no gate set can be simultaneously universal and transversal, as stated by the Eastin-Knill theorem \cite{Eastin2009}.

The standard approach for fault-tolerant quantum computation based on surface codes is to encode patches of physical qubits into logical qubits, then use a fermion-to-qubit encoding at the logical level to simulate fermionic systems \cite{Cody_Jones_2012}. In this setting, entangling gates are implemented using lattice surgery \cite{Horsman2012, Litinski_2019}. Alternatively, it is possible to perform fault-tolerant computations directly at the level of Majorana fermions collectively encoded through holes and twist-defects in a single surface code patch \cite{brown2017, Bravyi2002}. Here, no logical operators can be implemented transversally and instead defects/holes must be braided around each other, which involves performing measurements along the corresponding trajectories in the surface code lattice. 

For general quantum computations involving logical qubits, the first method is preferred since encoding defects at the boundaries of a surface code patch requires roughly three times less qubits
than encoding them in the bulk. Further, for braiding one has to ensure sufficient spatial separation between defects to prevent lowering the distance of the code which incurs an additional qubit overhead~\cite{Horsman2012}. On the other hand, additional ancillary surface code patches are required in order to locally implement logical entangling operations such as the CNOT gate through lattice surgery. Taking everything into account, the qubit requirements for lattice surgery turn out to be roughly 1.5-2 times lower than for braiding~\cite{Horsman2012}. 

If we instead consider computations on fermions, additional advantages favoring the braiding approach appear. First of all, the four Majorana operators which are created by a pair of twist-defects can only encode one qubit, while they can encode two fermionic modes, and as we have shown, there are ways of \emph{compressing} more fermionic modes into the surface code patch of a given physical qubit budget. Moreover, two-dimensional fermionic systems, such as the (spinful) Fermi-Hubbard model, can be encoded natively using twist-defects, such that local operators from the fermionic Hamiltonian remain local. When using fermion-to-qubit encodings at the logical level, on the other hand, this is not a priori guaranteed. For example, the Jordan-Wigner encoding leads to non-local high-weight operators for two-dimensional systems. Local encodings, such as the VC or DK preserve locality, but incur an additional qubit overhead of a factor $1.5-2$, as well as requiring deeper circuits to implement logical operations at the fermionic level. For a more detailed discussion of the relative advantages of different approaches to fault-tolerant fermionic simulation we defer the reader to Ref.~\cite{landahl2023} and references therein.

\begin{figure}
    \centering
    \includegraphics[width=\linewidth]{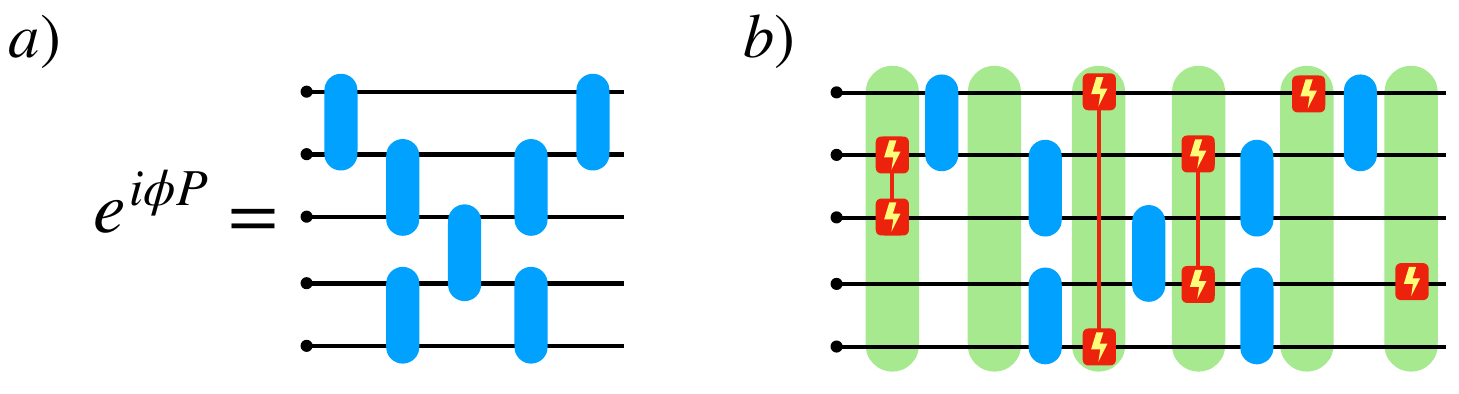}
    \caption{\textbf{Noise model}. a) Noiseless quantum circuit of an operator decomposition. b) Noisy implementation of this circuit with possible positions (shaded green areas) where errors (examples in red) can occur for the considered error model, limited to Pauli errors with weights up to two.}
    \label{fig:noise_model}
\end{figure}

All of the aforementioned techniques for fault-tolerant implementations of fermion-to-qubit encodings require a substantial overhead in physical qubits as well as deep circuits, which drastically restricts their applicability in the near-term era. Alternatively, one may consider using higher-distance fermion-to-qubit encodings for quantum error mitigation purposes, combined with allowing for the non-fault-tolerant implementation of logical operations. By increasing the code distance, it is possible to detect or correct an exponentially growing number of errors, but higher-weight operators and additional qubits will also introduce more noise into the system. Further, because of the non-fault-tolerant implementation of logical operators, small detectable errors can potentially quickly propagate through the circuit and transform into undetectable ones. To analyze this inherent trade-off, previous studies have focused on the comparison of encodings with various operator weights and distances \cite{chien2023}.

We focus on analyzing quantum error mitigation based on symmetry verification, which can be implemented by performing a parity-check after any logical operator \cite{Bonet_Monroig_2018}. If an error is detected, the shot is discarded and the experiment is re-run. This strategy is often referred to as post-selection or quantum error detection. While it is clear that the number of discarded shots increases with the code distance, we will assume that this overhead is practically tenable, and focus on the question of whether increasing the distance improves the accuracy of the final result, which is determined by the probability of undetected errors (those commuting with stabilizers) occurring within a given encoding. When scaling up the encoding, the weight of the errors that can be detected increases, but there are two effects that counteract this improvement: 1) increasing the number of gates and qubits leads to generally noisier circuits and 2) additional entangling operations may increase the weights of errors as they propagate through the circuit. 

Let us consider the error mitigation of a single logical operator $e^{i\phi P}$ with $P$ a Pauli string of weight $d$, as any improvements at this level will directly translate to the performance of the entire algorithm. In the best case, such as for vertex and transfer operators in the one-dimensional LE, the weights of all logical operators are equal to the distance. We therefore specifically focus on the implementation of a weight-$d$ vertex operator, and do not expect the results of the analysis to be affected by this particular choice of logical operator. We decompose the vertex operator via Clifford ladders using the XYZ decomposition \cite{Sriluckshmy2023} into a native gate set composed of operators of the form $e^{i\theta P_i}$, where $P_i$ is a single or two-qubit Pauli operator. This decomposition, depicted in Fig.~\ref{fig:noise_model}a, is composed of outer layers of Clifford operators, i.e. $\theta=\pi/4$ while the central gate encodes the angle of the rotation $\phi$, generally representing a non-Clifford gate. We consider a Pauli error model in which the noise is applied between layers of gates as depicted in Fig.~\ref{fig:noise_model}b.

Even though logical operators are generally not Clifford, Pauli errors can be transported from each layer to the end of the decomposition. A distance $d$ encoding will be able to detect any Pauli error with weight $w\leq d-1$. We neglect the possibility of multiple errors occurring simultaneously, since any practical quantum simulation will require the implementation of single logical operators with a reasonably high fidelity.

In this setup, the only errors that can lead to non-detectable (ND) errors are those in the layers before and after the central gate since no other errors can be transported to a weight $w>d-1$ error after conjugating the gates from the operator decomposition. Single-qubit errors cannot lead to ND errors when the central gate is a two-qubit gate, but two-qubit errors can if they connect two different leg operators of the decomposition. Every two-qubit error that will cause an ND error after transportation for distance $d$, will also create an ND error with the same probability for $d+1$. This is because all two-qubit Pauli errors causing ND faults in the decomposition at distance $d$ have one-to-one correspondence to such errors in a $d+1$ decomposition. Logical operators of distance $d+1$ then, are at least as likely to have ND errors as the distance $d$ ones. Below distance $d<4$, increasing $d$ to $d+1$ will also allow for new two-qubit errors to appear that create ND errors up to $d<4$. Since the number of ND errors does not improve with increasing distance but the amount of detectable errors worsens, low-distance encodings are expected to require less shots while proportionally detecting more or the same number of errors. 

Numerical comparisons for more QEM techniques and different encodings are given in Ref.~\cite{Papic2025} that show $\mathrm{LE}(2)$ to be the best candidate for a range of noise parameters.

Moving beyond error mitigation via error detection, fermionic encodings with sufficiently high distance $d\geq3$ also allow for the correction of errors. Following a similar approach to the the case of error detection, we now focus on the near-term implementation of partial error correction and study the probability of a non-correctable error occurring in the circuit implementing a vertex operator. For error correction, unless fault-tolerant executions are used, growing the distance is expected not to improve the performance, based on the same arguments used for the error mitigation case, and considering that only errors of weight up to $\floor{(d-1)/2}$ can be corrected.

\section{Conclusions}\label{sec:conclusions}
In this paper, we have introduced a number of families of fermion-to-qubit encodings, which can be grown to arbitrary code distances. This was achieved by building efficient low-weight fermion-to-qubit encodings and translating them into the language of topological defects on surface codes. This allowed us to grow the minimal weight of logical operators, and thus the distance, by spacing defects further apart.

We introduced the family of Ladder Encodings, which map one-dimensional fermionic lattice Hamiltonians into optimal weight Pauli operators for a given code distance, but are also adapted in this work to two-dimensional fermionic lattice Hamiltonians. Higher-connected Hamiltonians, like the electronic structure Hamiltonian, can be in principle also simulated via fSWAP routing on Ladder Encodings. Even though the LE$(d)$ can be constructed by following previous twist-defect strategies, its operators are more compact than previous distance-scalable encodings due to the more beneficial choice of Majorana pairs ordering. In particular, the 1D LE$(2)$ is optimal in the sense that the weight of vertex and transfer logical operators are equal to the distance of the encoding.

We applied our strategy for growing the distance to previously fixed-distance fermion-to-qubit encodings from literature, such as Verstaete-Cirac, Derby-Klassen and the hexagonal encoding. This strategy can be equally applied to other encodings by representing them at the level of Majorana operators, which can then be straightforwardly translated into topological defects within the surface code. Whilst we have mainly focused our attention on encoding twist-defects within the surface code, we have also shown a proof-of-principle construction on top of the 6.6.6. color code. One additional benefit of studying the surface code as opposed to other QEC codes, is that efficient techniques for unitary state preparation \cite{higgott2021optimal} and fault-tolerant implementations of quantum algorithms already exist \cite{brown2017,landahl2023}. Further work could include extending this approach to specific instances of other QEC codes, in particular for growing the distance of three-dimensional fermion-to-qubit encodings \cite{Derby2021}, or to fermion-to-qudit local encodings \cite{carobene2024} using qudit topological codes \cite{Anwar_2014}, such as the qudit surface code \cite{Bullock_2007}.

We have explored the usefulness of scaling the distance for performing noise mitigation based on post-selection and non-fault tolerant error correction strategies. We argued that higher distance encodings perform worse in this setting due to the impact of additional noise generated by deeper circuits of higher-weight operators. This means that fermion-to-qubit encodings aimed at near-term simulation should be mainly designed to reduce the weights of all required logical operators, which leads to shallower circuits. As a result, the most useful encodings from this work for near-term fermionic simulations are the $\mathrm{LE}(2)$ for one-dimensional systems (as shown in Ref.\cite{Papic2025}), $\mathrm{HX}(2)$ for two-dimensional systems of sufficient size since it allows for vertical hoppings to be implemented locally, or alternatively the PE$(3)$ for Hamiltonians with two spin species if the qubit connectivity is restricted or if the implementation of spin-flip terms is required. We note that the encodings identified in this paper are highly competitive, in terms of logical operator and stabilizer weights as well as their fermion-to-qubit ratios, compared to other state-of-the-art encodings with $d=3$ \cite{Simkovic2024}. For $d\geq4$ our encodings have similar logical operator weight compared to state-of-the-art, much lower stabilizer weights, yet at the price of significantly higher fermion-to-qubit ratios. This apparent trade-off between the qubit requirements and locality/simplicity of fermion-to-qubit encodings is in fact reminiscent of the very similar trade-off one faces between the surface code and quantum low-density parity check (qLDPC) codes. 

\section*{Acknowledgements}
The authors thank Francisco Revson Fernandes Pereira, Pedro Parrado-Rodr\'iguez and Martin Leib for useful discussions. This work was supported by the German Federal Ministry of Education and Research (BMBF), within the Research Program Quantum Systems, via the joint project QUBE (grant number 13N17152).

\bibliography{bibliography}

\begin{thebibliography}{49}%
\makeatletter
\providecommand \@ifxundefined [1]{%
 \@ifx{#1\undefined}
}%
\providecommand \@ifnum [1]{%
 \ifnum #1\expandafter \@firstoftwo
 \else \expandafter \@secondoftwo
 \fi
}%
\providecommand \@ifx [1]{%
 \ifx #1\expandafter \@firstoftwo
 \else \expandafter \@secondoftwo
 \fi
}%
\providecommand \natexlab [1]{#1}%
\providecommand \enquote  [1]{``#1''}%
\providecommand \bibnamefont  [1]{#1}%
\providecommand \bibfnamefont [1]{#1}%
\providecommand \citenamefont [1]{#1}%
\providecommand \href@noop [0]{\@secondoftwo}%
\providecommand \href [0]{\begingroup \@sanitize@url \@href}%
\providecommand \@href[1]{\@@startlink{#1}\@@href}%
\providecommand \@@href[1]{\endgroup#1\@@endlink}%
\providecommand \@sanitize@url [0]{\catcode `\\12\catcode `\$12\catcode `\&12\catcode `\#12\catcode `\^12\catcode `\_12\catcode `\%12\relax}%
\providecommand \@@startlink[1]{}%
\providecommand \@@endlink[0]{}%
\providecommand \url  [0]{\begingroup\@sanitize@url \@url }%
\providecommand \@url [1]{\endgroup\@href {#1}{\urlprefix }}%
\providecommand \urlprefix  [0]{URL }%
\providecommand \Eprint [0]{\href }%
\providecommand \doibase [0]{https://doi.org/}%
\providecommand \selectlanguage [0]{\@gobble}%
\providecommand \bibinfo  [0]{\@secondoftwo}%
\providecommand \bibfield  [0]{\@secondoftwo}%
\providecommand \translation [1]{[#1]}%
\providecommand \BibitemOpen [0]{}%
\providecommand \bibitemStop [0]{}%
\providecommand \bibitemNoStop [0]{.\EOS\space}%
\providecommand \EOS [0]{\spacefactor3000\relax}%
\providecommand \BibitemShut  [1]{\csname bibitem#1\endcsname}%
\let\auto@bib@innerbib\@empty
\bibitem [{\citenamefont {McArdle}\ \emph {et~al.}(2020)\citenamefont {McArdle}, \citenamefont {Endo}, \citenamefont {Aspuru-Guzik}, \citenamefont {Benjamin},\ and\ \citenamefont {Yuan}}]{McArdle_2020}%
  \BibitemOpen
  \bibfield  {author} {\bibinfo {author} {\bibfnamefont {S.}~\bibnamefont {McArdle}}, \bibinfo {author} {\bibfnamefont {S.}~\bibnamefont {Endo}}, \bibinfo {author} {\bibfnamefont {A.}~\bibnamefont {Aspuru-Guzik}}, \bibinfo {author} {\bibfnamefont {S.~C.}\ \bibnamefont {Benjamin}},\ and\ \bibinfo {author} {\bibfnamefont {X.}~\bibnamefont {Yuan}},\ }\bibfield  {title} {\bibinfo {title} {Quantum computational chemistry},\ }\bibfield  {journal} {\bibinfo  {journal} {Reviews of Modern Physics}\ }\textbf {\bibinfo {volume} {92}},\ \href {https://doi.org/10.1103/revmodphys.92.015003} {10.1103/revmodphys.92.015003} (\bibinfo {year} {2020})\BibitemShut {NoStop}%
\bibitem [{\citenamefont {Su}\ \emph {et~al.}(2021)\citenamefont {Su}, \citenamefont {Berry}, \citenamefont {Wiebe}, \citenamefont {Rubin},\ and\ \citenamefont {Babbush}}]{Su_2021}%
  \BibitemOpen
  \bibfield  {author} {\bibinfo {author} {\bibfnamefont {Y.}~\bibnamefont {Su}}, \bibinfo {author} {\bibfnamefont {D.~W.}\ \bibnamefont {Berry}}, \bibinfo {author} {\bibfnamefont {N.}~\bibnamefont {Wiebe}}, \bibinfo {author} {\bibfnamefont {N.}~\bibnamefont {Rubin}},\ and\ \bibinfo {author} {\bibfnamefont {R.}~\bibnamefont {Babbush}},\ }\bibfield  {title} {\bibinfo {title} {Fault-tolerant quantum simulations of chemistry in first quantization},\ }\bibfield  {journal} {\bibinfo  {journal} {PRX Quantum}\ }\textbf {\bibinfo {volume} {2}},\ \href {https://doi.org/10.1103/prxquantum.2.040332} {10.1103/prxquantum.2.040332} (\bibinfo {year} {2021})\BibitemShut {NoStop}%
\bibitem [{\citenamefont {Jordan}\ and\ \citenamefont {Wigner}(1928)}]{Jordan1928}%
  \BibitemOpen
  \bibfield  {author} {\bibinfo {author} {\bibfnamefont {P.}~\bibnamefont {Jordan}}\ and\ \bibinfo {author} {\bibfnamefont {E.}~\bibnamefont {Wigner}},\ }\bibfield  {title} {\bibinfo {title} {Über das paulische Äquivalenzverbot},\ }\href {https://doi.org/10.1007/bf01331938} {\bibfield  {journal} {\bibinfo  {journal} {Zeitschrift für Physik}\ }\textbf {\bibinfo {volume} {47}},\ \bibinfo {pages} {631} (\bibinfo {year} {1928})}\BibitemShut {NoStop}%
\bibitem [{\citenamefont {Kitaev}(1995)}]{Kitaev1995}%
  \BibitemOpen
  \bibfield  {author} {\bibinfo {author} {\bibfnamefont {A.~Y.}\ \bibnamefont {Kitaev}},\ }\href {https://doi.org/10.48550/ARXIV.QUANT-PH/9511026} {\bibinfo {title} {Quantum measurements and the {A}belian stabilizer problem}} (\bibinfo {year} {1995})\BibitemShut {NoStop}%
\bibitem [{\citenamefont {Friesner}(2005)}]{Friesner2005}%
  \BibitemOpen
  \bibfield  {author} {\bibinfo {author} {\bibfnamefont {R.~A.}\ \bibnamefont {Friesner}},\ }\bibfield  {title} {\bibinfo {title} {Ab initioquantum chemistry: Methodology and applications},\ }\href {https://doi.org/10.1073/pnas.0408036102} {\bibfield  {journal} {\bibinfo  {journal} {Proceedings of the National Academy of Sciences}\ }\textbf {\bibinfo {volume} {102}},\ \bibinfo {pages} {6648–6653} (\bibinfo {year} {2005})}\BibitemShut {NoStop}%
\bibitem [{\citenamefont {Guaita}(2025)}]{Guaita_2025}%
  \BibitemOpen
  \bibfield  {author} {\bibinfo {author} {\bibfnamefont {T.}~\bibnamefont {Guaita}},\ }\bibfield  {title} {\bibinfo {title} {On the locality of qubit encodings of local fermionic modes},\ }\href {https://doi.org/10.22331/q-2025-02-25-1644} {\bibfield  {journal} {\bibinfo  {journal} {Quantum}\ }\textbf {\bibinfo {volume} {9}},\ \bibinfo {pages} {1644} (\bibinfo {year} {2025})}\BibitemShut {NoStop}%
\bibitem [{\citenamefont {Chien}\ \emph {et~al.}(2023)\citenamefont {Chien}, \citenamefont {Setia}, \citenamefont {Bonet-Monroig}, \citenamefont {Steudtner},\ and\ \citenamefont {Whitfield}}]{chien2023}%
  \BibitemOpen
  \bibfield  {author} {\bibinfo {author} {\bibfnamefont {R.~W.}\ \bibnamefont {Chien}}, \bibinfo {author} {\bibfnamefont {K.}~\bibnamefont {Setia}}, \bibinfo {author} {\bibfnamefont {X.}~\bibnamefont {Bonet-Monroig}}, \bibinfo {author} {\bibfnamefont {M.}~\bibnamefont {Steudtner}},\ and\ \bibinfo {author} {\bibfnamefont {J.~D.}\ \bibnamefont {Whitfield}},\ }\href {https://arxiv.org/abs/2303.02270} {\bibinfo {title} {Simulating quantum error mitigation in fermionic encodings}} (\bibinfo {year} {2023}),\ \Eprint {https://arxiv.org/abs/2303.02270} {arXiv:2303.02270 [quant-ph]} \BibitemShut {NoStop}%
\bibitem [{\citenamefont {Landahl}\ and\ \citenamefont {Morrison}(2023)}]{landahl2023}%
  \BibitemOpen
  \bibfield  {author} {\bibinfo {author} {\bibfnamefont {A.~J.}\ \bibnamefont {Landahl}}\ and\ \bibinfo {author} {\bibfnamefont {B.~C.~A.}\ \bibnamefont {Morrison}},\ }\href {https://arxiv.org/abs/2110.10280} {\bibinfo {title} {Logical fermions for fault-tolerant quantum simulation}} (\bibinfo {year} {2023}),\ \Eprint {https://arxiv.org/abs/2110.10280} {arXiv:2110.10280 [quant-ph]} \BibitemShut {NoStop}%
\bibitem [{\citenamefont {Parella-Dilmé}\ \emph {et~al.}(2024)\citenamefont {Parella-Dilmé}, \citenamefont {Kottmann}, \citenamefont {Zambrano}, \citenamefont {Mortimer}, \citenamefont {Kottmann},\ and\ \citenamefont {Acín}}]{Parella2024}%
  \BibitemOpen
  \bibfield  {author} {\bibinfo {author} {\bibfnamefont {T.}~\bibnamefont {Parella-Dilmé}}, \bibinfo {author} {\bibfnamefont {K.}~\bibnamefont {Kottmann}}, \bibinfo {author} {\bibfnamefont {L.}~\bibnamefont {Zambrano}}, \bibinfo {author} {\bibfnamefont {L.}~\bibnamefont {Mortimer}}, \bibinfo {author} {\bibfnamefont {J.~S.}\ \bibnamefont {Kottmann}},\ and\ \bibinfo {author} {\bibfnamefont {A.}~\bibnamefont {Acín}},\ }\bibfield  {title} {\bibinfo {title} {Reducing entanglement with physically inspired fermion-to-qubit mappings},\ }\bibfield  {journal} {\bibinfo  {journal} {PRX Quantum}\ }\textbf {\bibinfo {volume} {5}},\ \href {https://doi.org/10.1103/prxquantum.5.030333} {10.1103/prxquantum.5.030333} (\bibinfo {year} {2024})\BibitemShut {NoStop}%
\bibitem [{\citenamefont {Jiang}\ \emph {et~al.}(2019)\citenamefont {Jiang}, \citenamefont {McClean}, \citenamefont {Babbush},\ and\ \citenamefont {Neven}}]{Jiang2019}%
  \BibitemOpen
  \bibfield  {author} {\bibinfo {author} {\bibfnamefont {Z.}~\bibnamefont {Jiang}}, \bibinfo {author} {\bibfnamefont {J.}~\bibnamefont {McClean}}, \bibinfo {author} {\bibfnamefont {R.}~\bibnamefont {Babbush}},\ and\ \bibinfo {author} {\bibfnamefont {H.}~\bibnamefont {Neven}},\ }\bibfield  {title} {\bibinfo {title} {Majorana loop stabilizer codes for error mitigation in fermionic quantum simulations},\ }\bibfield  {journal} {\bibinfo  {journal} {Physical Review Applied}\ }\textbf {\bibinfo {volume} {12}},\ \href {https://doi.org/10.1103/physrevapplied.12.064041} {10.1103/physrevapplied.12.064041} (\bibinfo {year} {2019})\BibitemShut {NoStop}%
\bibitem [{\citenamefont {Derby}\ \emph {et~al.}(2021)\citenamefont {Derby}, \citenamefont {Klassen}, \citenamefont {Bausch},\ and\ \citenamefont {Cubitt}}]{Derby2020}%
  \BibitemOpen
  \bibfield  {author} {\bibinfo {author} {\bibfnamefont {C.}~\bibnamefont {Derby}}, \bibinfo {author} {\bibfnamefont {J.}~\bibnamefont {Klassen}}, \bibinfo {author} {\bibfnamefont {J.}~\bibnamefont {Bausch}},\ and\ \bibinfo {author} {\bibfnamefont {T.}~\bibnamefont {Cubitt}},\ }\bibfield  {title} {\bibinfo {title} {Compact fermion to qubit mappings},\ }\bibfield  {journal} {\bibinfo  {journal} {Physical Review B}\ }\textbf {\bibinfo {volume} {104}},\ \href {https://doi.org/10.1103/physrevb.104.035118} {10.1103/physrevb.104.035118} (\bibinfo {year} {2021})\BibitemShut {NoStop}%
\bibitem [{\citenamefont {Verstraete}\ and\ \citenamefont {Cirac}(2005)}]{Verstraete2005}%
  \BibitemOpen
  \bibfield  {author} {\bibinfo {author} {\bibfnamefont {F.}~\bibnamefont {Verstraete}}\ and\ \bibinfo {author} {\bibfnamefont {J.~I.}\ \bibnamefont {Cirac}},\ }\bibfield  {title} {\bibinfo {title} {Mapping local {Hamiltonian}s of fermions to local {Hamiltonian}s of spins},\ }\href {https://doi.org/10.1088/1742-5468/2005/09/p09012} {\bibfield  {journal} {\bibinfo  {journal} {Journal of Statistical Mechanics: Theory and Experiment}\ }\textbf {\bibinfo {volume} {2005}},\ \bibinfo {pages} {P09012} (\bibinfo {year} {2005})}\BibitemShut {NoStop}%
\bibitem [{\citenamefont {Ball}(2005)}]{Ball2005}%
  \BibitemOpen
  \bibfield  {author} {\bibinfo {author} {\bibfnamefont {R.~C.}\ \bibnamefont {Ball}},\ }\bibfield  {title} {\bibinfo {title} {Fermions without fermion fields},\ }\bibfield  {journal} {\bibinfo  {journal} {Physical Review Letters}\ }\textbf {\bibinfo {volume} {95}},\ \href {https://doi.org/10.1103/physrevlett.95.176407} {10.1103/physrevlett.95.176407} (\bibinfo {year} {2005})\BibitemShut {NoStop}%
\bibitem [{\citenamefont {Chiew}\ and\ \citenamefont {Strelchuk}(2021)}]{Chiew2021}%
  \BibitemOpen
  \bibfield  {author} {\bibinfo {author} {\bibfnamefont {M.}~\bibnamefont {Chiew}}\ and\ \bibinfo {author} {\bibfnamefont {S.}~\bibnamefont {Strelchuk}},\ }\href {https://doi.org/10.48550/ARXIV.2110.12792} {\bibinfo {title} {Optimal fermion-qubit mappings}} (\bibinfo {year} {2021})\BibitemShut {NoStop}%
\bibitem [{\citenamefont {Miller}\ \emph {et~al.}(2023)\citenamefont {Miller}, \citenamefont {Zimborás}, \citenamefont {Knecht}, \citenamefont {Maniscalco},\ and\ \citenamefont {García-Pérez}}]{Miller_2023}%
  \BibitemOpen
  \bibfield  {author} {\bibinfo {author} {\bibfnamefont {A.}~\bibnamefont {Miller}}, \bibinfo {author} {\bibfnamefont {Z.}~\bibnamefont {Zimborás}}, \bibinfo {author} {\bibfnamefont {S.}~\bibnamefont {Knecht}}, \bibinfo {author} {\bibfnamefont {S.}~\bibnamefont {Maniscalco}},\ and\ \bibinfo {author} {\bibfnamefont {G.}~\bibnamefont {García-Pérez}},\ }\bibfield  {title} {\bibinfo {title} {Bonsai algorithm: Grow your own fermion-to-qubit mappings},\ }\bibfield  {journal} {\bibinfo  {journal} {PRX Quantum}\ }\textbf {\bibinfo {volume} {4}},\ \href {https://doi.org/10.1103/prxquantum.4.030314} {10.1103/prxquantum.4.030314} (\bibinfo {year} {2023})\BibitemShut {NoStop}%
\bibitem [{\citenamefont {Jiang}\ \emph {et~al.}(2020)\citenamefont {Jiang}, \citenamefont {Kalev}, \citenamefont {Mruczkiewicz},\ and\ \citenamefont {Neven}}]{Jiang2020}%
  \BibitemOpen
  \bibfield  {author} {\bibinfo {author} {\bibfnamefont {Z.}~\bibnamefont {Jiang}}, \bibinfo {author} {\bibfnamefont {A.}~\bibnamefont {Kalev}}, \bibinfo {author} {\bibfnamefont {W.}~\bibnamefont {Mruczkiewicz}},\ and\ \bibinfo {author} {\bibfnamefont {H.}~\bibnamefont {Neven}},\ }\bibfield  {title} {\bibinfo {title} {Optimal fermion-to-qubit mapping via ternary trees with applications to reduced quantum states learning},\ }\href {https://doi.org/10.22331/q-2020-06-04-276} {\bibfield  {journal} {\bibinfo  {journal} {Quantum}\ }\textbf {\bibinfo {volume} {4}},\ \bibinfo {pages} {276} (\bibinfo {year} {2020})}\BibitemShut {NoStop}%
\bibitem [{\citenamefont {Vlasov}(2022)}]{Vlasov_2022}%
  \BibitemOpen
  \bibfield  {author} {\bibinfo {author} {\bibfnamefont {A.~Y.}\ \bibnamefont {Vlasov}},\ }\bibfield  {title} {\bibinfo {title} {Clifford algebras, spin groups and qubit trees},\ }\href {https://doi.org/10.12743/quanta.v11i1.199} {\bibfield  {journal} {\bibinfo  {journal} {Quanta}\ }\textbf {\bibinfo {volume} {11}},\ \bibinfo {pages} {97–114} (\bibinfo {year} {2022})}\BibitemShut {NoStop}%
\bibitem [{\citenamefont {Setia}\ \emph {et~al.}(2019)\citenamefont {Setia}, \citenamefont {Bravyi}, \citenamefont {Mezzacapo},\ and\ \citenamefont {Whitfield}}]{Setia2019}%
  \BibitemOpen
  \bibfield  {author} {\bibinfo {author} {\bibfnamefont {K.}~\bibnamefont {Setia}}, \bibinfo {author} {\bibfnamefont {S.}~\bibnamefont {Bravyi}}, \bibinfo {author} {\bibfnamefont {A.}~\bibnamefont {Mezzacapo}},\ and\ \bibinfo {author} {\bibfnamefont {J.~D.}\ \bibnamefont {Whitfield}},\ }\bibfield  {title} {\bibinfo {title} {Superfast encodings for fermionic quantum simulation},\ }\bibfield  {journal} {\bibinfo  {journal} {Physical Review Research}\ }\textbf {\bibinfo {volume} {1}},\ \href {https://doi.org/10.1103/physrevresearch.1.033033} {10.1103/physrevresearch.1.033033} (\bibinfo {year} {2019})\BibitemShut {NoStop}%
\bibitem [{\citenamefont {Chen}\ \emph {et~al.}(2024)\citenamefont {Chen}, \citenamefont {Gorshkov},\ and\ \citenamefont {Xu}}]{Chen_2024}%
  \BibitemOpen
  \bibfield  {author} {\bibinfo {author} {\bibfnamefont {Y.-A.}\ \bibnamefont {Chen}}, \bibinfo {author} {\bibfnamefont {A.~V.}\ \bibnamefont {Gorshkov}},\ and\ \bibinfo {author} {\bibfnamefont {Y.}~\bibnamefont {Xu}},\ }\bibfield  {title} {\bibinfo {title} {Error-correcting codes for fermionic quantum simulation},\ }\bibfield  {journal} {\bibinfo  {journal} {SciPost Physics}\ }\textbf {\bibinfo {volume} {16}},\ \href {https://doi.org/10.21468/scipostphys.16.1.033} {10.21468/scipostphys.16.1.033} (\bibinfo {year} {2024})\BibitemShut {NoStop}%
\bibitem [{\citenamefont {Chien}\ and\ \citenamefont {Klassen}(2022)}]{Chien2022}%
  \BibitemOpen
  \bibfield  {author} {\bibinfo {author} {\bibfnamefont {R.~W.}\ \bibnamefont {Chien}}\ and\ \bibinfo {author} {\bibfnamefont {J.}~\bibnamefont {Klassen}},\ }\href {https://doi.org/10.48550/ARXIV.2210.05652} {\bibinfo {title} {Optimizing fermionic encodings for both {H}amiltonian and hardware}} (\bibinfo {year} {2022})\BibitemShut {NoStop}%
\bibitem [{\citenamefont {au2}\ \emph {et~al.}(2024)\citenamefont {au2}, \citenamefont {Leib},\ and\ \citenamefont {Pereira}}]{Simkovic2024}%
  \BibitemOpen
  \bibfield  {author} {\bibinfo {author} {\bibfnamefont {F.~S.~I.}\ \bibnamefont {au2}}, \bibinfo {author} {\bibfnamefont {M.}~\bibnamefont {Leib}},\ and\ \bibinfo {author} {\bibfnamefont {F.~R.~F.}\ \bibnamefont {Pereira}},\ }\href {https://arxiv.org/abs/2402.15386} {\bibinfo {title} {Low-weight high-distance error correcting fermionic encodings}} (\bibinfo {year} {2024}),\ \Eprint {https://arxiv.org/abs/2402.15386} {arXiv:2402.15386 [quant-ph]} \BibitemShut {NoStop}%
\bibitem [{\citenamefont {Setia}\ and\ \citenamefont {Whitfield}(2018)}]{Setia2018}%
  \BibitemOpen
  \bibfield  {author} {\bibinfo {author} {\bibfnamefont {K.}~\bibnamefont {Setia}}\ and\ \bibinfo {author} {\bibfnamefont {J.~D.}\ \bibnamefont {Whitfield}},\ }\bibfield  {title} {\bibinfo {title} {Bravyi-kitaev superfast simulation of electronic structure on a quantum computer},\ }\href {https://doi.org/10.1063/1.5019371} {\bibfield  {journal} {\bibinfo  {journal} {The Journal of Chemical Physics}\ }\textbf {\bibinfo {volume} {148}},\ \bibinfo {pages} {164104} (\bibinfo {year} {2018})}\BibitemShut {NoStop}%
\bibitem [{\citenamefont {Chen}\ and\ \citenamefont {Xu}(2023)}]{Chen_2023}%
  \BibitemOpen
  \bibfield  {author} {\bibinfo {author} {\bibfnamefont {Y.-A.}\ \bibnamefont {Chen}}\ and\ \bibinfo {author} {\bibfnamefont {Y.}~\bibnamefont {Xu}},\ }\bibfield  {title} {\bibinfo {title} {Equivalence between fermion-to-qubit mappings in two spatial dimensions},\ }\bibfield  {journal} {\bibinfo  {journal} {PRX Quantum}\ }\textbf {\bibinfo {volume} {4}},\ \href {https://doi.org/10.1103/prxquantum.4.010326} {10.1103/prxquantum.4.010326} (\bibinfo {year} {2023})\BibitemShut {NoStop}%
\bibitem [{\citenamefont {Algaba}\ \emph {et~al.}(2024)\citenamefont {Algaba}, \citenamefont {Sriluckshmy}, \citenamefont {Leib},\ and\ \citenamefont {Å imkovic IV}}]{Algaba_2024}%
  \BibitemOpen
  \bibfield  {author} {\bibinfo {author} {\bibfnamefont {M.~G.}\ \bibnamefont {Algaba}}, \bibinfo {author} {\bibfnamefont {P.~V.}\ \bibnamefont {Sriluckshmy}}, \bibinfo {author} {\bibfnamefont {M.}~\bibnamefont {Leib}},\ and\ \bibinfo {author} {\bibfnamefont {F.}~\bibnamefont {Å imkovic IV}},\ }\bibfield  {title} {\bibinfo {title} {Low-depth simulations of fermionic systems on square-grid quantum hardware},\ }\href {https://doi.org/10.22331/q-2024-04-30-1327} {\bibfield  {journal} {\bibinfo  {journal} {Quantum}\ }\textbf {\bibinfo {volume} {8}},\ \bibinfo {pages} {1327} (\bibinfo {year} {2024})}\BibitemShut {NoStop}%
\bibitem [{\citenamefont {Bravyi}\ and\ \citenamefont {Kitaev}(2002)}]{Bravyi2002}%
  \BibitemOpen
  \bibfield  {author} {\bibinfo {author} {\bibfnamefont {S.~B.}\ \bibnamefont {Bravyi}}\ and\ \bibinfo {author} {\bibfnamefont {A.~Y.}\ \bibnamefont {Kitaev}},\ }\bibfield  {title} {\bibinfo {title} {Fermionic quantum computation},\ }\href {https://doi.org/10.1006/aphy.2002.6254} {\bibfield  {journal} {\bibinfo  {journal} {Annals of Physics}\ }\textbf {\bibinfo {volume} {298}},\ \bibinfo {pages} {210} (\bibinfo {year} {2002})}\BibitemShut {NoStop}%
\bibitem [{\citenamefont {Bringewatt}\ and\ \citenamefont {Davoudi}(2023)}]{Bringewatt_2023}%
  \BibitemOpen
  \bibfield  {author} {\bibinfo {author} {\bibfnamefont {J.}~\bibnamefont {Bringewatt}}\ and\ \bibinfo {author} {\bibfnamefont {Z.}~\bibnamefont {Davoudi}},\ }\bibfield  {title} {\bibinfo {title} {Parallelization techniques for quantum simulation of fermionic systems},\ }\href {https://doi.org/10.22331/q-2023-04-13-975} {\bibfield  {journal} {\bibinfo  {journal} {Quantum}\ }\textbf {\bibinfo {volume} {7}},\ \bibinfo {pages} {975} (\bibinfo {year} {2023})}\BibitemShut {NoStop}%
\bibitem [{\citenamefont {Nielsen}\ and\ \citenamefont {Chuang}(2012)}]{Nielsen2012}%
  \BibitemOpen
  \bibfield  {author} {\bibinfo {author} {\bibfnamefont {M.~A.}\ \bibnamefont {Nielsen}}\ and\ \bibinfo {author} {\bibfnamefont {I.~L.}\ \bibnamefont {Chuang}},\ }\href {https://doi.org/10.1017/cbo9780511976667} {\emph {\bibinfo {title} {Quantum Computation and Quantum Information: 10th Anniversary Edition}}}\ (\bibinfo  {publisher} {Cambridge University Press},\ \bibinfo {year} {2012})\BibitemShut {NoStop}%
\bibitem [{\citenamefont {Litinski}\ and\ \citenamefont {von Oppen}(2018)}]{Litinski_2018}%
  \BibitemOpen
  \bibfield  {author} {\bibinfo {author} {\bibfnamefont {D.}~\bibnamefont {Litinski}}\ and\ \bibinfo {author} {\bibfnamefont {F.}~\bibnamefont {von Oppen}},\ }\bibfield  {title} {\bibinfo {title} {Quantum computing with majorana fermion codes},\ }\bibfield  {journal} {\bibinfo  {journal} {Physical Review B}\ }\textbf {\bibinfo {volume} {97}},\ \href {https://doi.org/10.1103/physrevb.97.205404} {10.1103/physrevb.97.205404} (\bibinfo {year} {2018})\BibitemShut {NoStop}%
\bibitem [{\citenamefont {Brown}\ \emph {et~al.}(2017)\citenamefont {Brown}, \citenamefont {Laubscher}, \citenamefont {Kesselring},\ and\ \citenamefont {Wootton}}]{brown2017}%
  \BibitemOpen
  \bibfield  {author} {\bibinfo {author} {\bibfnamefont {B.~J.}\ \bibnamefont {Brown}}, \bibinfo {author} {\bibfnamefont {K.}~\bibnamefont {Laubscher}}, \bibinfo {author} {\bibfnamefont {M.~S.}\ \bibnamefont {Kesselring}},\ and\ \bibinfo {author} {\bibfnamefont {J.~R.}\ \bibnamefont {Wootton}},\ }\bibfield  {title} {\bibinfo {title} {Poking holes and cutting corners to achieve clifford gates with the surface code},\ }\bibfield  {journal} {\bibinfo  {journal} {Physical Review X}\ }\textbf {\bibinfo {volume} {7}},\ \href {https://doi.org/10.1103/physrevx.7.021029} {10.1103/physrevx.7.021029} (\bibinfo {year} {2017})\BibitemShut {NoStop}%
\bibitem [{\citenamefont {Bombin}(2010)}]{bombin2010}%
  \BibitemOpen
  \bibfield  {author} {\bibinfo {author} {\bibfnamefont {H.}~\bibnamefont {Bombin}},\ }\bibfield  {title} {\bibinfo {title} {Topological order with a twist: Ising anyons from an abelian model},\ }\bibfield  {journal} {\bibinfo  {journal} {Physical Review Letters}\ }\textbf {\bibinfo {volume} {105}},\ \href {https://doi.org/10.1103/physrevlett.105.030403} {10.1103/physrevlett.105.030403} (\bibinfo {year} {2010})\BibitemShut {NoStop}%
\bibitem [{\citenamefont {Bombin}\ and\ \citenamefont {Martin-Delgado}(2006)}]{bombincolorcode2006}%
  \BibitemOpen
  \bibfield  {author} {\bibinfo {author} {\bibfnamefont {H.}~\bibnamefont {Bombin}}\ and\ \bibinfo {author} {\bibfnamefont {M.~A.}\ \bibnamefont {Martin-Delgado}},\ }\bibfield  {title} {\bibinfo {title} {Topological quantum distillation},\ }\bibfield  {journal} {\bibinfo  {journal} {Physical Review Letters}\ }\textbf {\bibinfo {volume} {97}},\ \href {https://doi.org/10.1103/physrevlett.97.180501} {10.1103/physrevlett.97.180501} (\bibinfo {year} {2006})\BibitemShut {NoStop}%
\bibitem [{\citenamefont {Landahl}\ \emph {et~al.}(2011)\citenamefont {Landahl}, \citenamefont {Anderson},\ and\ \citenamefont {Rice}}]{landahl2011faulttolerantquantumcomputingcolor}%
  \BibitemOpen
  \bibfield  {author} {\bibinfo {author} {\bibfnamefont {A.~J.}\ \bibnamefont {Landahl}}, \bibinfo {author} {\bibfnamefont {J.~T.}\ \bibnamefont {Anderson}},\ and\ \bibinfo {author} {\bibfnamefont {P.~R.}\ \bibnamefont {Rice}},\ }\href {https://arxiv.org/abs/1108.5738} {\bibinfo {title} {Fault-tolerant quantum computing with color codes}} (\bibinfo {year} {2011}),\ \Eprint {https://arxiv.org/abs/1108.5738} {arXiv:1108.5738 [quant-ph]} \BibitemShut {NoStop}%
\bibitem [{\citenamefont {Tiurev}\ \emph {et~al.}(2024)\citenamefont {Tiurev}, \citenamefont {Pesah}, \citenamefont {Derks}, \citenamefont {Roffe}, \citenamefont {Eisert}, \citenamefont {Kesselring},\ and\ \citenamefont {Reiner}}]{Tiurev_2024}%
  \BibitemOpen
  \bibfield  {author} {\bibinfo {author} {\bibfnamefont {K.}~\bibnamefont {Tiurev}}, \bibinfo {author} {\bibfnamefont {A.}~\bibnamefont {Pesah}}, \bibinfo {author} {\bibfnamefont {P.-J.~H.}\ \bibnamefont {Derks}}, \bibinfo {author} {\bibfnamefont {J.}~\bibnamefont {Roffe}}, \bibinfo {author} {\bibfnamefont {J.}~\bibnamefont {Eisert}}, \bibinfo {author} {\bibfnamefont {M.~S.}\ \bibnamefont {Kesselring}},\ and\ \bibinfo {author} {\bibfnamefont {J.-M.}\ \bibnamefont {Reiner}},\ }\bibfield  {title} {\bibinfo {title} {Domain wall color code},\ }\bibfield  {journal} {\bibinfo  {journal} {Physical Review Letters}\ }\textbf {\bibinfo {volume} {133}},\ \href {https://doi.org/10.1103/physrevlett.133.110601} {10.1103/physrevlett.133.110601} (\bibinfo {year} {2024})\BibitemShut {NoStop}%
\bibitem [{\citenamefont {Kesselring}\ \emph {et~al.}(2018)\citenamefont {Kesselring}, \citenamefont {Pastawski}, \citenamefont {Eisert},\ and\ \citenamefont {Brown}}]{Kesselring_2018}%
  \BibitemOpen
  \bibfield  {author} {\bibinfo {author} {\bibfnamefont {M.~S.}\ \bibnamefont {Kesselring}}, \bibinfo {author} {\bibfnamefont {F.}~\bibnamefont {Pastawski}}, \bibinfo {author} {\bibfnamefont {J.}~\bibnamefont {Eisert}},\ and\ \bibinfo {author} {\bibfnamefont {B.~J.}\ \bibnamefont {Brown}},\ }\bibfield  {title} {\bibinfo {title} {The boundaries and twist defects of the color code and their applications to topological quantum computation},\ }\href {https://doi.org/10.22331/q-2018-10-19-101} {\bibfield  {journal} {\bibinfo  {journal} {Quantum}\ }\textbf {\bibinfo {volume} {2}},\ \bibinfo {pages} {101} (\bibinfo {year} {2018})}\BibitemShut {NoStop}%
\bibitem [{\citenamefont {Kesselring}\ \emph {et~al.}(2024)\citenamefont {Kesselring}, \citenamefont {Magdalena de~la Fuente}, \citenamefont {Thomsen}, \citenamefont {Eisert}, \citenamefont {Bartlett},\ and\ \citenamefont {Brown}}]{Kesselring_2024}%
  \BibitemOpen
  \bibfield  {author} {\bibinfo {author} {\bibfnamefont {M.~S.}\ \bibnamefont {Kesselring}}, \bibinfo {author} {\bibfnamefont {J.~C.}\ \bibnamefont {Magdalena de~la Fuente}}, \bibinfo {author} {\bibfnamefont {F.}~\bibnamefont {Thomsen}}, \bibinfo {author} {\bibfnamefont {J.}~\bibnamefont {Eisert}}, \bibinfo {author} {\bibfnamefont {S.~D.}\ \bibnamefont {Bartlett}},\ and\ \bibinfo {author} {\bibfnamefont {B.~J.}\ \bibnamefont {Brown}},\ }\bibfield  {title} {\bibinfo {title} {Anyon condensation and the color code},\ }\bibfield  {journal} {\bibinfo  {journal} {PRX Quantum}\ }\textbf {\bibinfo {volume} {5}},\ \href {https://doi.org/10.1103/prxquantum.5.010342} {10.1103/prxquantum.5.010342} (\bibinfo {year} {2024})\BibitemShut {NoStop}%
\bibitem [{\citenamefont {Kivlichan}\ \emph {et~al.}(2018)\citenamefont {Kivlichan}, \citenamefont {McClean}, \citenamefont {Wiebe}, \citenamefont {Gidney}, \citenamefont {Aspuru-Guzik}, \citenamefont {Chan},\ and\ \citenamefont {Babbush}}]{Kivlichan2018}%
  \BibitemOpen
  \bibfield  {author} {\bibinfo {author} {\bibfnamefont {I.~D.}\ \bibnamefont {Kivlichan}}, \bibinfo {author} {\bibfnamefont {J.}~\bibnamefont {McClean}}, \bibinfo {author} {\bibfnamefont {N.}~\bibnamefont {Wiebe}}, \bibinfo {author} {\bibfnamefont {C.}~\bibnamefont {Gidney}}, \bibinfo {author} {\bibfnamefont {A.}~\bibnamefont {Aspuru-Guzik}}, \bibinfo {author} {\bibfnamefont {G.~K.-L.}\ \bibnamefont {Chan}},\ and\ \bibinfo {author} {\bibfnamefont {R.}~\bibnamefont {Babbush}},\ }\bibfield  {title} {\bibinfo {title} {Quantum simulation of electronic structure with linear depth and connectivity},\ }\bibfield  {journal} {\bibinfo  {journal} {Physical Review Letters}\ }\textbf {\bibinfo {volume} {120}},\ \href {https://doi.org/10.1103/physrevlett.120.110501} {10.1103/physrevlett.120.110501} (\bibinfo {year} {2018})\BibitemShut {NoStop}%
\bibitem [{\citenamefont {Projansky}\ \emph {et~al.}(2024)\citenamefont {Projansky}, \citenamefont {Heath},\ and\ \citenamefont {Whitfield}}]{Projansky2024}%
  \BibitemOpen
  \bibfield  {author} {\bibinfo {author} {\bibfnamefont {A.~M.}\ \bibnamefont {Projansky}}, \bibinfo {author} {\bibfnamefont {J.~T.}\ \bibnamefont {Heath}},\ and\ \bibinfo {author} {\bibfnamefont {J.~D.}\ \bibnamefont {Whitfield}},\ }\bibfield  {title} {\bibinfo {title} {Entanglement spectrum of matchgate circuits with universal and non-universal resources},\ }\href {https://doi.org/10.22331/q-2024-08-07-1432} {\bibfield  {journal} {\bibinfo  {journal} {Quantum}\ }\textbf {\bibinfo {volume} {8}},\ \bibinfo {pages} {1432} (\bibinfo {year} {2024})}\BibitemShut {NoStop}%
\bibitem [{\citenamefont {Eastin}\ and\ \citenamefont {Knill}(2009)}]{Eastin2009}%
  \BibitemOpen
  \bibfield  {author} {\bibinfo {author} {\bibfnamefont {B.}~\bibnamefont {Eastin}}\ and\ \bibinfo {author} {\bibfnamefont {E.}~\bibnamefont {Knill}},\ }\bibfield  {title} {\bibinfo {title} {Restrictions on transversal encoded quantum gate sets},\ }\bibfield  {journal} {\bibinfo  {journal} {Physical Review Letters}\ }\textbf {\bibinfo {volume} {102}},\ \href {https://doi.org/10.1103/physrevlett.102.110502} {10.1103/physrevlett.102.110502} (\bibinfo {year} {2009})\BibitemShut {NoStop}%
\bibitem [{\citenamefont {Cody~Jones}\ \emph {et~al.}(2012)\citenamefont {Cody~Jones}, \citenamefont {Whitfield}, \citenamefont {McMahon}, \citenamefont {Yung}, \citenamefont {Meter}, \citenamefont {Aspuru-Guzik},\ and\ \citenamefont {Yamamoto}}]{Cody_Jones_2012}%
  \BibitemOpen
  \bibfield  {author} {\bibinfo {author} {\bibfnamefont {N.}~\bibnamefont {Cody~Jones}}, \bibinfo {author} {\bibfnamefont {J.~D.}\ \bibnamefont {Whitfield}}, \bibinfo {author} {\bibfnamefont {P.~L.}\ \bibnamefont {McMahon}}, \bibinfo {author} {\bibfnamefont {M.-H.}\ \bibnamefont {Yung}}, \bibinfo {author} {\bibfnamefont {R.~V.}\ \bibnamefont {Meter}}, \bibinfo {author} {\bibfnamefont {A.}~\bibnamefont {Aspuru-Guzik}},\ and\ \bibinfo {author} {\bibfnamefont {Y.}~\bibnamefont {Yamamoto}},\ }\bibfield  {title} {\bibinfo {title} {Faster quantum chemistry simulation on fault-tolerant quantum computers},\ }\href {https://doi.org/10.1088/1367-2630/14/11/115023} {\bibfield  {journal} {\bibinfo  {journal} {New Journal of Physics}\ }\textbf {\bibinfo {volume} {14}},\ \bibinfo {pages} {115023} (\bibinfo {year} {2012})}\BibitemShut {NoStop}%
\bibitem [{\citenamefont {Horsman}\ \emph {et~al.}(2012)\citenamefont {Horsman}, \citenamefont {Fowler}, \citenamefont {Devitt},\ and\ \citenamefont {Meter}}]{Horsman2012}%
  \BibitemOpen
  \bibfield  {author} {\bibinfo {author} {\bibfnamefont {D.}~\bibnamefont {Horsman}}, \bibinfo {author} {\bibfnamefont {A.~G.}\ \bibnamefont {Fowler}}, \bibinfo {author} {\bibfnamefont {S.}~\bibnamefont {Devitt}},\ and\ \bibinfo {author} {\bibfnamefont {R.~V.}\ \bibnamefont {Meter}},\ }\bibfield  {title} {\bibinfo {title} {Surface code quantum computing by lattice surgery},\ }\href {https://doi.org/10.1088/1367-2630/14/12/123011} {\bibfield  {journal} {\bibinfo  {journal} {New Journal of Physics}\ }\textbf {\bibinfo {volume} {14}},\ \bibinfo {pages} {123011} (\bibinfo {year} {2012})}\BibitemShut {NoStop}%
\bibitem [{\citenamefont {Litinski}(2019)}]{Litinski_2019}%
  \BibitemOpen
  \bibfield  {author} {\bibinfo {author} {\bibfnamefont {D.}~\bibnamefont {Litinski}},\ }\bibfield  {title} {\bibinfo {title} {A game of surface codes: Large-scale quantum computing with lattice surgery},\ }\href {https://doi.org/10.22331/q-2019-03-05-128} {\bibfield  {journal} {\bibinfo  {journal} {Quantum}\ }\textbf {\bibinfo {volume} {3}},\ \bibinfo {pages} {128} (\bibinfo {year} {2019})}\BibitemShut {NoStop}%
\bibitem [{\citenamefont {Bonet-Monroig}\ \emph {et~al.}(2018)\citenamefont {Bonet-Monroig}, \citenamefont {Sagastizabal}, \citenamefont {Singh},\ and\ \citenamefont {O'Brien}}]{Bonet_Monroig_2018}%
  \BibitemOpen
  \bibfield  {author} {\bibinfo {author} {\bibfnamefont {X.}~\bibnamefont {Bonet-Monroig}}, \bibinfo {author} {\bibfnamefont {R.}~\bibnamefont {Sagastizabal}}, \bibinfo {author} {\bibfnamefont {M.}~\bibnamefont {Singh}},\ and\ \bibinfo {author} {\bibfnamefont {T.~E.}\ \bibnamefont {O'Brien}},\ }\bibfield  {title} {\bibinfo {title} {Low-cost error mitigation by symmetry verification},\ }\bibfield  {journal} {\bibinfo  {journal} {Physical Review A}\ }\textbf {\bibinfo {volume} {98}},\ \href {https://doi.org/10.1103/physreva.98.062339} {10.1103/physreva.98.062339} (\bibinfo {year} {2018})\BibitemShut {NoStop}%
\bibitem [{\citenamefont {Sriluckshmy}\ \emph {et~al.}(2023)\citenamefont {Sriluckshmy}, \citenamefont {Pina-Canelles}, \citenamefont {Ponce}, \citenamefont {Algaba}, \citenamefont {IV},\ and\ \citenamefont {Leib}}]{Sriluckshmy2023}%
  \BibitemOpen
  \bibfield  {author} {\bibinfo {author} {\bibfnamefont {P.~V.}\ \bibnamefont {Sriluckshmy}}, \bibinfo {author} {\bibfnamefont {V.}~\bibnamefont {Pina-Canelles}}, \bibinfo {author} {\bibfnamefont {M.}~\bibnamefont {Ponce}}, \bibinfo {author} {\bibfnamefont {M.~G.}\ \bibnamefont {Algaba}}, \bibinfo {author} {\bibfnamefont {F.~Å.}\ \bibnamefont {IV}},\ and\ \bibinfo {author} {\bibfnamefont {M.}~\bibnamefont {Leib}},\ }\bibfield  {title} {\bibinfo {title} {Optimal, hardware native decomposition of parameterized multi-qubit pauli gates},\ }\href {https://doi.org/10.1088/2058-9565/acfa20} {\bibfield  {journal} {\bibinfo  {journal} {Quantum Science and Technology}\ }\textbf {\bibinfo {volume} {8}},\ \bibinfo {pages} {045029} (\bibinfo {year} {2023})}\BibitemShut {NoStop}%
\bibitem [{\citenamefont {Papi\v{c}}\ \emph {et~al.}(2025)\citenamefont {Papi\v{c}}, \citenamefont {Godinez-Ramirez}, \citenamefont {Ines~de Vega}, \citenamefont {\v{S}imkovic IV},\ and\ \citenamefont {Calzona}}]{Papic2025}%
  \BibitemOpen
  \bibfield  {author} {\bibinfo {author} {\bibfnamefont {M.}~\bibnamefont {Papi\v{c}}}, \bibinfo {author} {\bibfnamefont {E.}~\bibnamefont {Godinez-Ramirez}}, \bibinfo {author} {\bibfnamefont {A.~A.}\ \bibnamefont {Ines~de Vega}}, \bibinfo {author} {\bibfnamefont {F.}~\bibnamefont {\v{S}imkovic IV}},\ and\ \bibinfo {author} {\bibfnamefont {A.}~\bibnamefont {Calzona}},\ }\href@noop {} {\bibinfo {title} {In preparation}} (\bibinfo {year} {2025})\BibitemShut {NoStop}%
\bibitem [{\citenamefont {Higgott}\ \emph {et~al.}(2021)\citenamefont {Higgott}, \citenamefont {Wilson}, \citenamefont {Hefford}, \citenamefont {Dborin}, \citenamefont {Hanif}, \citenamefont {Burton},\ and\ \citenamefont {Browne}}]{higgott2021optimal}%
  \BibitemOpen
  \bibfield  {author} {\bibinfo {author} {\bibfnamefont {O.}~\bibnamefont {Higgott}}, \bibinfo {author} {\bibfnamefont {M.}~\bibnamefont {Wilson}}, \bibinfo {author} {\bibfnamefont {J.}~\bibnamefont {Hefford}}, \bibinfo {author} {\bibfnamefont {J.}~\bibnamefont {Dborin}}, \bibinfo {author} {\bibfnamefont {F.}~\bibnamefont {Hanif}}, \bibinfo {author} {\bibfnamefont {S.}~\bibnamefont {Burton}},\ and\ \bibinfo {author} {\bibfnamefont {D.~E.}\ \bibnamefont {Browne}},\ }\bibfield  {title} {\bibinfo {title} {Optimal local unitary encoding circuits for the surface code},\ }\href {https://doi.org/10.22331/q-2021-08-05-517} {\bibfield  {journal} {\bibinfo  {journal} {Quantum}\ }\textbf {\bibinfo {volume} {5}},\ \bibinfo {pages} {517} (\bibinfo {year} {2021})}\BibitemShut {NoStop}%
\bibitem [{\citenamefont {Derby}\ and\ \citenamefont {Klassen}(2021)}]{Derby2021}%
  \BibitemOpen
  \bibfield  {author} {\bibinfo {author} {\bibfnamefont {C.}~\bibnamefont {Derby}}\ and\ \bibinfo {author} {\bibfnamefont {J.}~\bibnamefont {Klassen}},\ }\href {https://doi.org/10.48550/arxiv.2101.10735} {\bibinfo {title} {A compact fermion to qubit mapping part 2: Alternative lattice geometries}} (\bibinfo {year} {2021})\BibitemShut {NoStop}%
\bibitem [{\citenamefont {Carobene}\ \emph {et~al.}(2024)\citenamefont {Carobene}, \citenamefont {Barison}, \citenamefont {Giachero},\ and\ \citenamefont {Nys}}]{carobene2024}%
  \BibitemOpen
  \bibfield  {author} {\bibinfo {author} {\bibfnamefont {R.}~\bibnamefont {Carobene}}, \bibinfo {author} {\bibfnamefont {S.}~\bibnamefont {Barison}}, \bibinfo {author} {\bibfnamefont {A.}~\bibnamefont {Giachero}},\ and\ \bibinfo {author} {\bibfnamefont {J.}~\bibnamefont {Nys}},\ }\href {https://arxiv.org/abs/2412.05616} {\bibinfo {title} {Local fermion-to-qudit mappings}} (\bibinfo {year} {2024}),\ \Eprint {https://arxiv.org/abs/2412.05616} {arXiv:2412.05616 [quant-ph]} \BibitemShut {NoStop}%
\bibitem [{\citenamefont {Anwar}\ \emph {et~al.}(2014)\citenamefont {Anwar}, \citenamefont {Brown}, \citenamefont {Campbell},\ and\ \citenamefont {Browne}}]{Anwar_2014}%
  \BibitemOpen
  \bibfield  {author} {\bibinfo {author} {\bibfnamefont {H.}~\bibnamefont {Anwar}}, \bibinfo {author} {\bibfnamefont {B.~J.}\ \bibnamefont {Brown}}, \bibinfo {author} {\bibfnamefont {E.~T.}\ \bibnamefont {Campbell}},\ and\ \bibinfo {author} {\bibfnamefont {D.~E.}\ \bibnamefont {Browne}},\ }\bibfield  {title} {\bibinfo {title} {Fast decoders for qudit topological codes},\ }\href {https://doi.org/10.1088/1367-2630/16/6/063038} {\bibfield  {journal} {\bibinfo  {journal} {New Journal of Physics}\ }\textbf {\bibinfo {volume} {16}},\ \bibinfo {pages} {063038} (\bibinfo {year} {2014})}\BibitemShut {NoStop}%
\bibitem [{\citenamefont {Bullock}\ and\ \citenamefont {Brennen}(2007)}]{Bullock_2007}%
  \BibitemOpen
  \bibfield  {author} {\bibinfo {author} {\bibfnamefont {S.~S.}\ \bibnamefont {Bullock}}\ and\ \bibinfo {author} {\bibfnamefont {G.~K.}\ \bibnamefont {Brennen}},\ }\bibfield  {title} {\bibinfo {title} {Qudit surface codes and gauge theory with finite cyclic groups},\ }\href {https://doi.org/10.1088/1751-8113/40/13/013} {\bibfield  {journal} {\bibinfo  {journal} {Journal of Physics A: Mathematical and Theoretical}\ }\textbf {\bibinfo {volume} {40}},\ \bibinfo {pages} {3481–3505} (\bibinfo {year} {2007})}\BibitemShut {NoStop}%
\end{thebibliography}%

\end{document}